\documentclass[hidelinks,aps,rmp,twocolumn,superscriptaddress,10pt]{revtex4-1} 

\usepackage{amsmath,amssymb}
\usepackage{mathrsfs}
\usepackage{epstopdf}
\usepackage{graphicx}
\usepackage{bm,color,bbm}
\usepackage{natbib}
\usepackage{hyperref}
\usepackage{caption}
\usepackage{subcaption}

\newcommand{\nn}{\nonumber}

\newcommand{\ket}[1]{ |{#1} \rangle}

\begin{document}

\preprint{}

\title{Introduction to the Transverse Spatial Correlations in Spontaneous Parametric Down-Conversion through the Biphoton Birth Zone}


\author{James Schneeloch}
\email{jfizzix@gmail.com}
\affiliation{Department of Physics and Astronomy, University of Rochester, Rochester, New York 14627, USA}
\affiliation{Center for Coherence and Quantum Optics, University of Rochester, Rochester, New York 14627, USA}
\affiliation{Air Force Research Laboratory, Information Directorate, Rome, New York, 13441, USA}

\author{John C. Howell}
\email{howell@pas.rochester.edu}
\affiliation{Department of Physics and Astronomy, University of Rochester, Rochester, New York 14627, USA}
\affiliation{Center for Coherence and Quantum Optics, University of Rochester, Rochester, New York 14627, USA}


\date{\today}

\begin{abstract}
As a tutorial to the spatial aspects of Spontaneous Parametric Downconversion (SPDC), we present a detailed first-principles derivation of the transverse correlation width of photon pairs in degenerate collinear SPDC. This width defines the size of a biphoton birth zone, the region where the signal and idler photons are likely to be found when conditioning on the position of the destroyed pump photon. Along the way, we discuss the quantum-optical calculation of the amplitude for the SPDC process, as well as its simplified form for nearly collinear degenerate phase matching. Following this, we show how this biphoton amplitude can be approximated with a Double-Gaussian wavefunction, and give a brief discussion of the measurement statistics (and subsequent convenience) of such Double-Gaussian wavefunctions. Next, we use this approximation to get a simplified estimation of the transverse correlation width, and compare it to more accurate calculations as well as experimental results. We then conclude with a discussion of the concept of a biphoton birth zone, using it to develop intuition for the tradeoff between the first-order spatial coherence and bipohoton correlations in SPDC.
\end{abstract}


\maketitle
\tableofcontents

\section{Introduction}
In continuous-variable quantum information, there are many experiments using entangled photon pairs generated by spontaneous parametric downconversion (SPDC) \footnote{There are many dozens (if not hundreds) of experimental papers either using or exploring the spatial entanglement between photon pairs from SPDC, but some papers representative of the scope of research are: \cite{Reid2009, BroughamQKD2012, TascaQC2011, Leach2012, Schneeloch2012, edgar2012imaging, PhysRevA.86.010101, Howell2004, Walborn2011, HowlandPRX2013, AliKhan_ETent_PRA2006, walbornSPDC, bennink_PhysRevLett.92.033601, Barreiro2008, PhysRevA.88.023848}}. In short, SPDC is a $\chi^{(2)}$-nonlinear optical process occurring in birefringent crystals \footnote{In order to produce SPDC, one does not necessarily need a \emph{birefringent} crystal, but this is a popular way to ensure a constant phase relationship (also known as phase matching) between the pump photon, and the signal/idler photon pair.} where high energy ``pump'' photons are converted into pairs of low energy ``signal'' and ``idler'' photons. In particular, the pump field interacts coherently with the electromagnetic quantum vacuum via a nonlinear medium in such a way that as an individual event, a pump photon is destroyed, and two daughter photons (signal and idler) are created (this event happening many times). As this process is a parametric process (i.e., one by definition in which the initial and final states of the crystal are the same), the total energy and total momentum of the field must each be conserved. Because of this, the energies and momenta of the daughter photons are highly correlated, and their joint quantum state is highly entangled. These highly entangled photon pairs may be used for any number of purposes, ranging from fundamental tests of quantum mechanics, to almost any application requiring (two-party) quantum entanglement. 

In this tutorial, we discuss a particularly convenient and common variety of SPDC used in quantum optics experiments. In particular, we consider illuminating a nonlinear crystal with a collimated continuous-wave pump beam, and filtering the downcoverted light to collect only those photon pairs with frequencies nearly equal to each other (each being about half of the pump frequency). This degenerate collinear SPDC process is amenable to many approximations, especially considering that most optical experiments are done in the paraxial regime, where all measurements are taken relatively close to the optic axis (allowing many small-angle approximations). With this sort of experimental setup in mind, we discuss the theoretical treatment of such entangled photon pairs (from first principles) in sufficient detail so as to inform the understanding and curiosity of anyone seeking to discuss or undertake such experiments \footnote{For more extensive treatments of spontaneous parametric downconversion, we recommend the Ph.D. theses of Lijun Wang \cite{wang1992investigation}, Warren Grice \cite{Gricethesis}, and Paul Kwiat \cite{kwiatthesis}, as well as the Physics Reports article by S.P. Walborn \emph{et~al.}\cite{walbornSPDC}.}. Indeed as we discuss all the necessary concepts preceding each approximation, much of this discussion will be useful in understanding non-collinear and non-degenerate SPDC as well.

The rest of this paper is laid out as follows. In Section 2, we discuss the derivation of the quantum biphoton field state in SPDC, as discussed in \cite{HongMandelSPDCPRA1985}, and \cite{mandel1995optical}. In addition to this, we point out what factors contribute not only to the shape of the biphoton wavefunction (defined later), but also to the magnitude of the amplitude for the biphoton generation to take place. This is important, as it determines the overall likelihood of downconversion events, and gives important details to look for in new materials in the hopes of creating brighter sources of entangled photon pairs. In Section 3, we simplify the biphoton wavefunction for the case of degenerate, collinear SPDC in the paraxial regime using the results in \cite{MonkenSPDCPRA1998}. We also use geometrical arguments to explain the approximations allowed in the paraxial regime. In Section 4, we show how to further approximate this approximate biphoton wavefunction as a Double-Gaussian (as seen in \cite{LawEberly2004} and \cite{Fedorov2009}), as the multivariate Gaussian density is well studied, and is easier to work with. In doing so, we give derivations of common statistical parameters of the Double-Gaussian wavefunction, showing its convenience in multiple applications. In Section 5, we provide a calculation of the transverse correlation width, defined as the standard deviation of the transverse distance between the signal and idler photons' positions at the time of their creation. In Section 6, we explore the utility of the transverse correlation width, and introduce the concepts of the biphoton birth zone, and of the birth zone number as a measure of biphoton correlation. We conclude by using the birth zone number to gain a qualitative understanding of the tradeoff between the first-order spatial coherence and the measurable correlations between photon pairs in the downconverted fields.

\section{Foundation: The Quantum - Optical Calculation of the Biphoton state in SPDC}
The procedure to quantize the electromagnetic field as it is used in quantum optics \cite{mandel1995optical,loudon2000quantum} (as opposed to quantum field theory), is to: decompose the electromagnetic field into a sum over (cavity) modes; find Hamilton's equations of motion for each field mode; and assign to the classically conjugate variables (generalized coordinates and momenta), quantum-mechanically conjugate obervables, whose commutator is $i\hbar$. From these field observables, one can obtain a Hamiltonian operator describing the evolution of the quantum electromagnetic field, and in so doing, describe the evolution of any quantum-optical system. 

SPDC is a $\chi^{(2)}$-nonlinear process. To describe it \cite{mandel1995optical}, we begin with the classical Hamiltonian of the electromagnetic field;
\begin{equation}
\mathcal{H}_{EM}= \frac{1}{2}\int d^{3}r\;(\vec{\mathbf{D}}\cdot \vec{\mathbf{E}} + \vec{\mathbf{B}}\cdot\vec{\mathbf{H}}),
\end{equation}
where $\vec{\mathbf{D}}=\epsilon_{0}\vec{\mathbf{E}}+\vec{\mathbf{P}}$. Since the electric field amplitude of the incident light on a nonlinear medium is usually substantially smaller than the electric field strength binding the atoms in a material together, we can express the polarization field $\vec{\mathbf{P}}$ as a power series in the electric field strength \cite{boyd2007nonlinear}, so that
\begin{equation}
\vec{\mathbf{P}}= \epsilon_{0}\big[\chi^{(1)}\vec{\mathbf{E}} + \chi^{(2)}(\vec{\mathbf{E}})^{2} + \chi^{(3)}(\vec{\mathbf{E}})^{3}+...\big].
\end{equation}
Since the nonlinear interaction beyond second order is considered here to not appreciably affect the polarization, the classical Hamiltonian for the electromagnetic field can be broken up into two terms, one linear, and one nonlinear;
\begin{equation}
\mathcal{H}_{EM}= \mathcal{H}_{L} + \mathcal{H}_{NL},
\end{equation}
where,
\begin{equation}
\mathcal{H}_{NL}=\frac{1}{2}\epsilon_{o}\int d^{3}r\;\tilde{\chi}^{(2)}_{ijl}E_{i}(\vec{\mathbf{r}},t)E_{j}(\vec{\mathbf{r}},t)E_{l}(\vec{\mathbf{r}},t).
\end{equation}
Next, since the nonlinear susceptibility $\tilde{\chi}^{2}_{ijl}$ depends on pump, signal, and idler frequencies\footnote{At this point we would like to point out that we use the Einstein summation convention for $\tilde{\chi}^{(2)}_{ijl}E_{i}(\vec{\mathbf{r}},t)E_{j}(\vec{\mathbf{r}},t)E_{l}(\vec{\mathbf{r}},t)$.}, each of which are determined by their respective wave numbers, the nonlinear Hamiltonian is better broken down into its frequency components:
\begin{align}
\mathcal{H}_{NL}&=\frac{1}{2(\sqrt{2\pi})^{3}}\epsilon_{o}\int d^{3}r\!\!\!\sum_{\vec{\mathbf{k}}_{p},\vec{\mathbf{k}}_{1},\vec{\mathbf{k}}_{2}}\!\!\big[\tilde{\chi}^{(2)}_{ijl}(\omega(\vec{\mathbf{k}}_{p}),\omega(\vec{\mathbf{k}}_{1}),\omega(\vec{\mathbf{k}}_{2}))\nn\\
&\times E_{i}(\omega(\vec{\mathbf{k}}_{p}))E_{j}(\omega(\vec{\mathbf{k}}_{1}))E_{l}(\omega(\vec{\mathbf{k}}_{2}))\big],
\end{align}
where subscripts $1$ and $2$, are understood to refer to signal and idler modes, respectively.

To condense this paper, we note that when the field quantization is carried out, our electric field functions $E(\vec{\mathbf{r}},t)$ are replaced by the field observables $\hat{E}(\vec{\mathbf{r}},t)$, which separate into a sum of positive and negative frequency contributions $\hat{E}^{+}(\vec{\mathbf{r}},t)$, and $\hat{E}^{-}(\vec{\mathbf{r}},t)$, where
\begin{equation}
\hat{E}^{+}(\vec{\mathbf{r}},t)=\frac{1}{V^{\frac{1}{2}}}\sum_{\vec{\mathbf{k}},s}i\sqrt{\frac{\hbar \omega(\vec{\mathbf{k}})}{2 \epsilon_{0}}}\hat{a}_{\vec{\mathbf{k}},s}(t)\;\vec{\mathbf{\epsilon}}_{\vec{\mathbf{k}},s}\;e^{i \vec{\mathbf{k}}\cdot \vec{\mathbf{r}}},
\end{equation}
and $\hat{E}^{-}(\vec{\mathbf{r}},t)$ is the hermitian conjugate of $\hat{E}^{+}(\vec{\mathbf{r}},t)$. Here, $s$ is an index indicating component of polarization, $\vec{\mathbf{\epsilon}}$ is a unit polarization vector, and $\hat{a}_{\vec{\mathbf{k}},s}(t)$ is the photon annihilation operator at time $t$. In addition, $V$ is the quantization volume \footnote{The quantization volume is the volume of the hypothetical cavity containing the modes of the electromagnetic field. For simplicity, the cavity is taken to be rectangular, so the sum over modes is straightforward using boundary conditions in Cartesian coordinates. To get an accurate representation of the electromagnetic field in free space, we may take the quantization volume to be arbitrarily large.}, which in the standard quantization procedure, would be the volume of a cavity which can be taken to approach infinity for the free-space case.

With the electric field observables thus defined, we can obtain the quantum Hamiltonian of the electromagnetic field:
\begin{align}\label{QHgeneral}
\hat{H} &= \sum_{\vec{\mathbf{k}},s} \hbar \omega(\vec{\mathbf{k}})\hat{a}^{\dagger}_{\vec{\mathbf{k}},s}\hat{a}_{\vec{\mathbf{k}},s}\nn\\
&+\frac{1}{2}\epsilon_{o}\int d^{3}r\;\tilde{\chi}^{(2)}_{ijl}\hat{E}_{i}(\vec{\mathbf{r}},t)\hat{E}_{j}(\vec{\mathbf{r}},t)\hat{E}_{l}(\vec{\mathbf{r}},t),
\end{align}
where the first term is the linear contribution to the Hamiltonian. Though this Hamiltonian looks relatively simple, the field operator $\hat{E}(\vec{\mathbf{r}},t)=\hat{E}^{+}(\vec{\mathbf{r}},t)+\hat{E}^{-}(\vec{\mathbf{r}},t)$, so that the integral in \eqref{QHgeneral} actually contains eight terms. These terms correspond to all different $\chi^{2}$ processes (e.g., sum-frequency generation, difference-frequency generation, optical rectification, etc.), each of which has its own probability amplitude of occurring. However, given that we have a single input field (i.e., the pump field), and start with no photons in either of the signal and idler fields, the only energy-conserving contributions to the Hamiltonian (i.e., the only significant contributions \footnote{The reason the non-energy-conserving terms in \eqref{QHgeneral} can be neglected is due to the rotating wave approximation. In calculating the amplitude for the downconversion process, and converting to the interaction picture, all other contributions to this amplitude will have complex exponentials oscillating much faster than $\Delta\omega\equiv \omega_{1}+\omega_{2}-\omega_{p}$. Since each of these contributions (oscillating at frequency $\omega$) when integrated give amplitudes proportional to $\text{sinc}\big(\frac{\omega T}{2}\big)$, and the propagation time $T$ through the crystal is fixed, these Sinc functions become negligibly small for large $\omega$. Since $\Delta\omega$ is small for nearly degenerate SPDC, the energy-conserving contribution dominates over the non-conserving contributions.}) are transitions (forward and backward) where pump photons are annihilated, and signal-idler photon pairs are created.

Our first approximation (beyond what was done to get \eqref{QHgeneral} to begin with) is that the pump beam is bright enough to be treated classically, and that the pump intensity is not significantly diminished due to downconversion events. This ``undepleted pump" approximation, along with keeping only the energy-conserving terms, gives us the simplified Hamiltonian:
\begin{equation}
\hat{H} = H_{L}+\frac{1}{2}\epsilon_{o}\int d^{3}r\;\big(\tilde{\chi}^{(2)}_{ijl}(\vec{\mathbf{r}})E_{i}(\vec{\mathbf{r}},t)\hat{E}^{-}_{j}(\vec{\mathbf{r}},t)\hat{E}^{-}_{l}(\vec{\mathbf{r}},t) + h.c.\big),
\end{equation}
which we then expand in the modes of the signal, and idler fields;
\begin{align}
\hat{H} &= H_{L}\nn\\
&+\frac{1}{2}\epsilon_{o}\int d^{3}r\;\bigg(\frac{-1}{V}\sum_{\vec{\mathbf{k}}_{1},s_{1}}\sum_{\vec{\mathbf{k}}_{2},s_{2}}\tilde{\chi}^{(2)}_{ijl}(\vec{\mathbf{r}};\omega(\vec{\mathbf{k}_{p}}),\omega(\vec{\mathbf{k}_{1}}),\omega(\vec{\mathbf{k}_{2}}))\nn\\
&\times \sqrt{\frac{\hbar^{2} \omega(\vec{\mathbf{k}}_{1})\omega(\vec{\mathbf{k}}_{2})}{4 \epsilon_{0}^{2}}} e^{-i (\vec{\mathbf{k}}_{1}+\vec{\mathbf{k}}_{2})\cdot \vec{\mathbf{r}}} E_{i}(\vec{\mathbf{r}},t)\nn\\
&\times \hat{a}^{\dagger}_{\vec{\mathbf{k}}_{1},s_{1}}(t)\hat{a}^{\dagger}_{\vec{\mathbf{k}}_{2},s_{2}}(t)\;(\vec{\mathbf{\epsilon}}_{\vec{\mathbf{k}}_{1},s_{1}})_{j}(\vec{\mathbf{\epsilon}}_{\vec{\mathbf{k}}_{2},s_{2}})_{l}+ h.c.\bigg).
\end{align}
Note that here and throughout this paper, $h.c.$ stands for hermitian conjugate.

Next, we assume the pump is sufficiently narrowband, so that we can, to a good approximation, separate out the time dependence of the pump field as a complex exponential of frequency $\omega_{p}$. In addition, we assume the pump field to be sufficiently well-collimated so that, to a good approximation, we can also separate out the longitudinal dependence of the pump field \footnote{For a reference that examines in detail how the downconverted light is affected by the pump spatial profile, we recommend the reference \cite{PittmanPhysRevA.53.2804}. For a reference that treats SPDC with short pump pulses (as opposed to continuous-wave), see \cite{KellerPhysRevA.56.1534}.}. At this point, we define the transverse momenta $\vec{\mathbf{q}}_{p}$, $\vec{\mathbf{q}}_{1}$, and $\vec{\mathbf{q}}_{2}$,as the projections of the pump wave vector $\vec{\mathbf{k}}_{p}$, the signal wave vector $\vec{\mathbf{k}}_{1}$, and the idler wave vector $\vec{\mathbf{k}}_{2}$, onto the plane transverse to the optic axis respectively. We also define $k_{pz}$, $k_{1z}$, and $k_{2z}$, as the longitudinal components of the corresponding wave vectors. With this in mind, we express the pump field as an integral over plane waves:
\begin{equation}
E_{i}(\vec{\mathbf{r}},t)=\frac{1}{2\pi}\int d^{2}q_{p}\;\tilde{E}_{i}(\vec{\mathbf{q}}_{p},t)e^{i (\vec{\mathbf{q}}_{p}\cdot\vec{\mathbf{r}})}e^{i( k_{zp}z -\omega_{p}t)}.
\end{equation}

By separating out the transverse components of the wave vectors, we make the Hamiltonian easier to simplify in later steps. As one additional simplification, we define the pump polarization vector $\vec{\epsilon}_{\vec{\mathbf{k}}_{p}}$, so that $E_{i}(\vec{\mathbf{q}}_{p},t)=E(\vec{\mathbf{q}}_{p},t)(\vec{\epsilon}_{\vec{\mathbf{k}}_{p}})_{i}$. With the transverse components separated out, and the narrowband pump approximation made, the Hamiltonian takes the form:
\begin{align}
&\hat{H}= H_{L}\nn\\
&+\frac{1}{4\pi}\epsilon_{o}\!\!\int d^{3}r d^{2}q_{p}\bigg(\frac{-1}{V}\sum_{\vec{\mathbf{k}}_{1},s_{1}}\sum_{\vec{\mathbf{k}}_{2},s_{2}}\tilde{\chi}^{(2)}_{ijl}(\vec{\mathbf{r}};\omega(\vec{\mathbf{k}_{p}}),\omega(\vec{\mathbf{k}_{1}}),\omega(\vec{\mathbf{k}_{2}}))\nn\\
&\times (\vec{\mathbf{\epsilon}}_{\vec{\mathbf{k}}_{1},s_{1}})_{j}(\vec{\mathbf{\epsilon}}_{\vec{\mathbf{k}}_{2},s_{2}})_{l}(\vec{\epsilon}_{\vec{\mathbf{k}}_{p}})_{i}\sqrt{\frac{\hbar^{2} \omega(\vec{\mathbf{k}}_{1})\omega(\vec{\mathbf{k}}_{2})}{4 \epsilon_{0}^{2}}}\nn\\
&\times e^{-i (\Delta\vec{\mathbf{q}})\cdot \vec{\mathbf{r}}}e^{-i\Delta k_{z}z}e^{-i\omega_{p}t} \tilde{E}_{i}(\vec{\mathbf{q}}_{p},t)\hat{a}^{\dagger}_{\vec{\mathbf{k}}_{1},s_{1}}(t)\hat{a}^{\dagger}_{\vec{\mathbf{k}}_{2},s_{2}}(t)\;+ h.c.\bigg),
\end{align}
where we define $\Delta\vec{\mathbf{q}}\equiv \vec{\mathbf{q}}_{1}+\vec{\mathbf{q}}_{2}-\vec{\mathbf{q}}_{p}$, and $\Delta k_{z}\equiv k_{1z}+k_{2z}-k_{pz}$.

In most experimental setups (including the one we consider here), the nonlinear crystal is a simple rectangular prism, centered at $\vec{\mathbf{r}}=0$, and with side lengths $L_{x}$, $L_{y}$, and $L_{z}$. Here, we assume the crystal is isotropic, so that $\chi^{(2)}_{ijl}$ does not depend on $\vec{\mathbf{r}}$. To simplify the subsequent calculations, we assume the crystal to be embedded in a linear optical medium of the same index of refraction to avoid dealing with multiple reflections. Alternatively, we could assume the crystal has an anti-reflective coating to the same effect. We can then carry out the integral over the spatial coordinates (from $-\frac{L}{2}$ to $\frac{L}{2}$ in each direction) (such that $d^{3}r=dxdydz$), to get the Hamiltonian:
\begin{align}
&\hat{H}= H_{L}\nn\\
&+\frac{1}{4\pi}\epsilon_{o}\!\!\int\!d^{2}q_{p}\bigg[\frac{-L_{x}L_{y}L_{z}}{V}\!\!\sum_{\vec{\mathbf{k}}_{1},s_{1}}\!\sum_{\vec{\mathbf{k}}_{2},s_{2}}\!\!\big[\tilde{\chi}^{(2)}_{ijl}(\omega(\vec{\mathbf{k}_{p}}),\omega(\vec{\mathbf{k}_{1}}),\omega(\vec{\mathbf{k}_{2}}))\nn\\
&\times (\vec{\epsilon}_{\vec{\mathbf{k}}_{p}})_{i}(\vec{\mathbf{\epsilon}}_{\vec{\mathbf{k}}_{1},s_{1}})_{j}(\vec{\mathbf{\epsilon}}_{\vec{\mathbf{k}}_{2},s_{2}})_{l}\big]\sqrt{\frac{\hbar^{2} \omega(\vec{\mathbf{k}}_{1})\omega(\vec{\mathbf{k}}_{2})}{4 \epsilon_{0}^{2}}}\nn\\
&\times\text{sinc}\bigg(\!\frac{\Delta q_{x} L_{x}}{2}\!\bigg)\text{sinc}\bigg(\!\frac{\Delta q_{y} L_{y}}{2}\!\bigg)\text{sinc}\bigg(\!\frac{\Delta k_{z} L_{z}}{2}\!\bigg)e^{-i\omega_{p}t} \tilde{E}(\vec{\mathbf{q}}_{p},t)\nn\\
&\times\hat{a}^{\dagger}_{\vec{\mathbf{k}}_{1},s_{1}}(t)\hat{a}^{\dagger}_{\vec{\mathbf{k}}_{2},s_{2}}(t)+ h.c.\bigg].
\end{align}
Note that the Sinc function, $\text{sinc}(x)$, is defined here as $\sin(x)/x$.

To obtain the state of the downconverted fields, one can readily use first-order time-dependent perturbation theory. To see why this is, we can compare the nonlinear classical Hamiltonian to the linear Hamiltonian using typical experimental parameters of a pump field intensity of $1mW/mm^{2}$, and signal/idler intensities of about $1 pW/mm^{2}$. As such, the nonlinear contribution to the total Hamiltonian is indeed very small relative to the linear part, and the consequent results we obtain from these first order calculations should be quite accurate. Though this is the treatment we discuss, alternative higher order and non-perturbative derivations of the quantum state of down-converted light are also useful in examining the photon number statistics of down converted light, particularly when a sufficiently intense pump beam makes it significantly probable that multiple pairs will be generated at once through the simultaneous absorption of multiple pump photons. Indeed, the general two photon state is described as a multimode squeezed vacuum state, whose photon number statistics have been shown experimentally (and theoretically) to be such that the number of pairs created in a given time interval is approximately Poisson-distributed \cite{PhysRevLett.101.053601, christ2011}.

Using first-order time-dependent perturbation theory, the state of the signal and idler fields in the interaction picture can be computed as follows:
\begin{equation}
\ket{\Psi(t)}\approx \bigg(1 -\frac{i}{\hbar}\int_{0}^{t}dt' H_{NL}(t')\bigg)\ket{\Psi(0)}.
\end{equation}
Note that in the interaction picture, operators evolve according to the unperturbed Hamiltonian, so that $\hat{a}^{\dagger}(t)=\hat{a}^{\dagger}(0)e^{i\omega t}$. Here, the initial state of the signal and idler fields $\ket{\Psi(0)}$ is given to be the vacuum state $\ket{0_{1},0_{2}}$, which means that the hermitian conjugate (with its lowering operators) will not contribute to the state of the downconverted photon fields. 

Before we calculate the state of the downconverted photon fields (up to a normalization factor), we make use of some simplifying assumptions. First, we assume that the polarizations of the downconverted photons are fixed, so that we can neglect the sums over $s_{1}$ and $s_{2}$, effectively making the sum over one value. With this, the sum over the components of the nonlinear susceptibility is proportional to the value $d_{\text{eff}}\equiv\frac{1}{2}\chi^{(2)}_{\text{eff}}$, which is the effective experimentally determined coefficient for the nonlinear interaction. Second, we assume that the nonlinear crystal is much larger than the optical wavelengths considered here, so that the sums over $\vec{\mathbf{k}}_{1}$ and $\vec{\mathbf{k}}_{2}$ can be replaced by integrals in the following way:
\begin{equation}
\lim_{V\rightarrow\infty}\frac{1}{V}\sum_{\vec{\mathbf{k}}_{1},s_{1}}=\frac{1}{(2\pi)^{3}}\sum_{s_{1}}\int d^{3}k_{1}.
\end{equation}
With these simplifications, we can express the nonlinear Hamiltonian in such a way that both the sums are replaced by integrals, while still accurately reflecting the relative likelihood of downconversion events;
\begin{align}
H_{NL}&\approx C_{NL}d_{eff}\iint d^{3}k_{1}d^{3}k_{2}\;\sqrt{\omega(\vec{\mathbf{k}}_{1})\omega(\vec{\mathbf{k}}_{2})}\nn\\
&\times \int d^{2}q_{p}\; \bigg[\prod_{m=1}^{3}\text{sinc}\bigg(\frac{\Delta k_{m}L_{m}}{2}\bigg)\bigg]\tilde{E}(\vec{\mathbf{q}}_{p},t)e^{i \Delta\omega t}\nn\\
&\times \hat{a}^{\dagger}(\vec{\mathbf{k}}_{1})\hat{a}^{\dagger}(\vec{\mathbf{k}}_{2}),
\end{align}
where $m=\{1,2,3\}=\{x,y,z\}$, and $\Delta\omega \equiv \omega_{1}+\omega_{2}-\omega_{p}$, and $C_{NL}$ is a constant.  Note that here, $\Delta k_{x}=\Delta q_{x}$ and $\Delta k_{y}=\Delta q_{y}$, to condense notation. 

With one additional assumption, that the slowly-varying pump amplitude (excluding $e^{i\omega_{p}t}$) is essentially constant over the time light takes to propagate through the crystal, the integral over this nonlinear Hamiltonian becomes an integral of a constant times $e^{i\Delta\omega t}$. With this integral, we get our first look at the state of the downconverted field exiting the crystal:
\begin{align}
&\ket{\Psi_{SPDC}}\approx C_{0}\ket{0_{1},0_{2}}\nn\\
&+ C_{1} d_{eff} \sqrt{I_{p}T^{2}}\iint d^{3}k_{1}d^{3}k_{2}\;\Phi(\vec{\mathbf{k}}_{1},\vec{\mathbf{k}}_{2})\nn\\
&\times \sqrt{\omega(\vec{\mathbf{k}}_{1})\omega(\vec{\mathbf{k}}_{2})} e^{\frac{i \Delta\omega T}{2}}\text{sinc}\bigg(\frac{\Delta\omega T}{2}\bigg)\hat{a}^{\dagger}(\vec{\mathbf{k}}_{1})\hat{a}^{\dagger}(\vec{\mathbf{k}}_{2})\ket{0_{1},0_{2}}\label{biphotStat1}.
\end{align}
Here, $T$ is the time it takes light to travel through the crystal; $I_{p}$ is the intensity of the pump beam; $\ket{0_{1},0_{2}}$ is the (Fock) vacuum state with zero photons in the signal mode and zero photons in the idler mode, and;
\begin{align}\label{BPWF}
\Phi(\vec{\mathbf{k}}_{1},\vec{\mathbf{k}}_{2})&\equiv \int d^{2}q_{p}\; \bigg[\prod_{m=1}^{3}\text{sinc}\bigg(\frac{\Delta k_{m}L_{m}}{2}\bigg)\bigg]\nu(\vec{\mathbf{q}}_{p}),\nn\\
\end{align}
is, up to a normalization constant, the biphoton wavefunction in momentum space (where $\nu (\vec{\mathbf{q}}_{p})$ is the normalized pump amplitude spectrum). To see how this works, we note that the biphoton probability amplitude can be expressed as $\langle 0_{1},0_{2}|\hat{a}(\vec{\mathbf{k}}_{1})\hat{a}(\vec{\mathbf{k}}_{2})|\Psi_{SPDC}\rangle$. When we normalize this probability amplitude, by integrating its magnitude square over all values of $\vec{\mathbf{k}}_{1}$, and $\vec{\mathbf{k}}_{2}$, and setting this integral equal to unity, the resulting normalized probability amplitude has the necessary properties (for our purposes) of a biphoton wavefunction \footnote{Though it is debatable whether it is correct to speak of a biphoton wavefunction since expectation values are in fact carried out with $\ket{\Psi_{SPDC}}$, and $\Phi(\vec{\mathbf{k}}_{1}, \vec{\mathbf{k}}_{2})$ does not evolve according to the Schr{\"o}dinger equation ($\ket{\Psi_{SPDC}}$ does, though), $\Phi(\vec{\mathbf{k}}_{1}, \vec{\mathbf{k}}_{2})$ is a square-integrable function in a two- particle joint Hilbert space that accurately describes the relative measurement statistics of the biphotons.}. With approximations to be made in the next section, only $\Phi(\vec{\mathbf{k}}_{1},\vec{\mathbf{k}}_{2})$ will govern the transverse momentum probability distribution of the biphoton field\footnote{Note that the biphoton wavefunction \eqref{BPWF} is expressed as an integral over the rectangular crystal shape. For those interested in a derivation of the integral giving the biphoton wavefunction for a generalized crystal shape, see \cite{saldanha_AJP_2013}}. 

The factors preceding the biphoton wavefunction  are still important to understand because (with some algebra) they contribute to the rate of downconversion events $(R_{SPDC})$. In particular \footnote{Note that this downconversion rate comes from the approximation of near-perfect energy conservation: i.e., $\Delta\omega T<<1$. \cite{Helt:12}}:
\begin{equation}
R_{SPDC}\propto d_{eff}^{2} P_{p} L_{z}^{2},
\end{equation}
where $P_{p}$ is the pump power (in Watts). We also note from equation \eqref{biphotStat1} that the rate of downconversion events is also proportional to $\text{sinc}^{2}(\Delta\omega T/2)$, though this factor is essentially unity for the nearly degenerate frequencies of the signal and idler photons considered here. This proportionality also follows from more rigorous calculations of the rate of downconversion events \cite{HongMandelSPDCPRA1985,Kleinman_PR1968}, though only in the approximation where the minuscule signal/idler fields don't appreciably contribute to the likelihood of downconversion events. In those more rigorous calculations, the conversion efficiency (biphotons made per incident pump photon) is of the order $10^{-8}$, which again shows just how weak these signal/idler fields are relative to the pump field \footnote{Including the collection/coupling efficiencies in many quantum-optical experiments, the measured conversion efficiency is closer to $10^{-10}$.}, and why first-order perturbation theory is sufficient to get a reasonably accurate representation of the state of the downconverted fields. We also note that although beam size doesn't affect the global rate of downconversion events, it does affect the fraction of those downconversion events that are likely to be counted by a detector near the optic axis \footnote{The rate of downconversion events yielding biphotons propagating close to the optic axis increases with a smaller beam size, but only to a point. For a good summary, see \cite{Ling_PhysRevA.77.043834}. For a more detailed discussion on how focusing affects the fraction of downconverted light propagating near the optic axis, see \cite{LjunggrenPhysRevA.72.062301}. For a more rigorous discussion of how the rate of downconversion events (i.e., the signal/idler power) changes with the crystal length, see \cite{loudon2000quantum}.}. Even so, these factors are useful to know when selecting a crystal as a source of entangled photon pairs. For example, with a constant power pump beam, a longer crystal will be a brighter source of photon pairs. However, there is a tradeoff; the degree of correlation between the signal and idler photons decreases with increasing crystal length (as we shall show).

\section{Approximation for degenerate collinear SPDC}
To obtain a relatively simple expression for the biphoton field in SPDC, we have made multiple (though reasonable) simplifying assumptions. We have assumed that the pump is narrowband and collimated so that it is nearly monochromatic, while also having a momentum spectrum whose longitudinal components dominate over its transverse components. We next assumed that the pump is bright enough to be treated classically, but not so bright that the perturbation series approximation to the nonlinear polarization breaks down. In addition, we assumed that we need not consider multiple reflections, and that the crystal is large compared to an optical wavelength so that sums over spatial modes may be replaced by integrals. We have also assumed that the pump is bright enough that it is not attenuated appreciably due to downconversion events. 

Now, we consider the experimental case where we place frequency filters over photon detectors, so that we may only examine downconversion events which are degenerate (where $\omega_{1}=\omega_{2}$),and perfectly energy-conserving ($\Delta\omega=0$). In this case, along with all the previous assumptions made, we define a new constant of normalization $\tilde{C}_{1}$ (absorbing factors outside the integrals), and obtain the following simplified expression for the state of the downconverted field as seen in \cite{MonkenSPDCPRA1998};
\begin{align}
\ket{\Psi_{SPDC}}&\approx C_{0}\ket{0_{1},0_{2}}\\
 +&\tilde{C}_{1}\iint d^{3}k_{1}d^{3}k_{2}\;\Phi(\vec{\mathbf{k}}_{1},\vec{\mathbf{k}}_{2})\hat{a}^{\dagger}(\vec{\mathbf{k}}_{1})\hat{a}^{\dagger}(\vec{\mathbf{k}}_{2})\ket{0_{1},0_{2}}\nn.
\end{align}
Here, the biphoton wavefunction $\Phi(\vec{\mathbf{k}}_{1},\vec{\mathbf{k}}_{2})$ is as defined previously \eqref{BPWF}.
Next, we use the fact that the transverse dimensions of the crystal are much larger than the pump wavelength to carry out the integral over the transverse pump momentum.
\begin{align}
\Phi&(\vec{\mathbf{k}}_{1},\vec{\mathbf{k}}_{2})= \text{sinc}\bigg(\frac{\Delta k_{z}L_{z}}{2}\bigg)\nn\\
\times&\int d^{2}q_{p}\; \bigg[\text{sinc}\bigg(\frac{\Delta k_{x}L_{x}}{2}\bigg)\text{sinc}\bigg(\frac{\Delta k_{y}L_{y}}{2}\bigg)\bigg]\nu(k_{px},k_{py}),
\end{align}
The significant contributions of the Sinc functions to the integral will come from when, for example, $\Delta k_{x}<\frac{\pi k_{p}}{2\Omega}$, where $\Omega$ is the ratio of the width of the crystal $L_{x}$ to the pump wavelength $\lambda_{p}$. Where the crystal is much wider than a pump wavelength, $\Omega$ is large, and we see the Sinc function will only contribute significantly when $\Delta k_{x}$ is only a very small fraction of $k_{p}$. Thus, with a renormalization, the sincs act like delta functions, setting $\vec{\mathbf{q}}_{p}=\vec{\mathbf{q}}_{1}+\vec{\mathbf{q}}_{2}$, and giving us the biphoton wavefunction:
\begin{equation}
\Phi(\vec{\mathbf{k}}_{1},\vec{\mathbf{k}}_{2})= \mathcal{N}\text{sinc}\bigg(\frac{\Delta k_{z}L_{z}}{2}\bigg)\nu(\vec{\mathbf{q}}_{1}+\vec{\mathbf{q}}_{2}),
\end{equation}
where $\mathcal{N}$ is a normalization constant.

\begin{figure}[t]
\centerline{\includegraphics[width=\columnwidth]{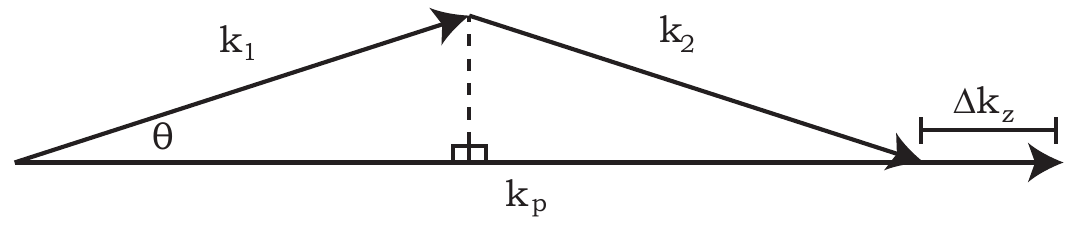}}
\caption{Diagram of the relationship between the pump, signal, and idler momenta, in standard (not necessarily collinear) phase matching.}
\end{figure} 

Since most experiments are done in the paraxial regime, we use such approximations to get the Sinc-Gaussian biphoton wavefunction, ubiquitous in the literature. With the previous assumptions already made, we point out that in degenerate, collinear SPDC, $|\vec{\mathbf{k}}_{1}|=|\vec{\mathbf{k}}_{2}|=|\vec{\mathbf{k}}_{p}|/2$, which we redefine as $k_{1}$,$k_{2}$, and $k_{p}/2$, to simplify notation. In addition, since the transverse pump momentum is essentially equal to the sum of the transverse signal and idler momenta, the three vectors can be readily drawn on a plane, as seen in Fig.~1.

Let $\theta$ be the angle between the pump momentum $\vec{\mathbf{k}}_{p}$  and the signal or idler momentum vectors $\vec{\mathbf{k}}_{1}$ and $\vec{\mathbf{k}}_{2}$. This angle $\theta$ is small enough that we may use the small-angle approximation to find an expression for $\Delta k_{z}$ in terms of easier-to-measure quantities. Using the conservation of each component of the total momentum, we get the following equations:
\begin{align}\label{angles}
k_{p}&=(k_{1}+k_{2})\cos(\theta) - \Delta k_{z},\\
\frac{|\vec{\mathbf{q}}_{1}-\vec{\mathbf{q}}_{2}|}{2}&=k_{1} \sin(\theta).
\end{align}
Using the small-angle approximation, and substituting one equation into the other, we find:
\begin{equation}
\Delta k_{z}\approx -\frac{|\vec{\mathbf{q}}_{1}-\vec{\mathbf{q}}_{2}|^{2}}{2 k_{p}}.
\end{equation}
Finally, when we assume the transverse pump momentum profile is a Gaussian,
\begin{equation}
\nu(\vec{\mathbf{q}}_{p})=\bigg(\frac{2 \sigma_{p}^{2}}{\pi}\bigg)^{\frac{1}{4}}e^{-\sigma_{p}^{2}|\vec{\mathbf{q}}_{p}|^{2}},
\end{equation} 
with $\sigma_{p}$ being the pump radius in position space \footnote{The pump radius $\sigma_{p}$ in position space is defined as the standard deviation of $\frac{x_{1}+x_{2}}{2}$. This is justified by noting that the pump radius in momentum space is explicitly given by the standard deviation of $(k_{1x}+k_{2x})$, and using the properties of Fourier transformed Gaussian wavefunctions.}, we renormalize, and find the biphoton wavefunction to be:
\begin{equation}\label{SGmomWavf}
\Phi(\vec{\mathbf{k}}_{1},\vec{\mathbf{k}}_{2})= \mathcal{N}\text{sinc}\bigg(\frac{L_{z}\lambda_{p}}{8\pi n_{p}}|\vec{\mathbf{q}}_{1}-\vec{\mathbf{q}}_{2}|^{2}\bigg)e^{-\sigma_{p}^{2}|\vec{\mathbf{q}}_{1}+\vec{\mathbf{q}}_{2}|^{2}},
\end{equation}
where the minus sign in the argument of the sinc function is eliminated since the sinc function is an even function. Interestingly, we can use the radius to the first zero of the Sinc function along with \eqref{angles} to derive a simple formula for the half-angle divergence of the degenerate collinear SPDC light:
\begin{equation}
\theta_{SPDC}\approx\sqrt{\frac{2\lambda_{p}}{L_{z}}}.
\end{equation}
In experiments where no filtering takes place to isolate the degenerate portion of the SPDC light, this angle will be larger since the non-degenerate frequencies of SPDC light have a wider, ring-shaped distribution.

To obtain a transverse correlation width from this biphoton wavefunction, we need to transform it to position space. Fortunately, this biphoton wavefunction is approximately \footnote{For small values of $x$ and $y$, $\text{sinc}(x+y)\sim \text{sinc}(x)\text{sinc}(y)$. For typical experimental parameters, the argument of the Sinc function is of the order $10^{-3}$, even for transverse momenta as large as the pump momentum. With the paraxial approximation, the transverse momenta are much smaller than the pump momentum, and so the arguments of the Sinc functions are very small indeed.} separable (subject to our paraxial approximation) into horizontal and vertical wavefunctions (i.e., into a product of functions, one dependent on only x-coordinates, and the other dependent on only y-coordinates). In addition, we can find an orthogonal set of coordinates in terms of sums and differences of momenta that allows us to transform this wavefunction by transforming the Sinc function and Gaussian individually. While transforming the Gaussian is extremely straightforward, transforming the concurrent Sinc-based function is more challenging, owing to that it is a Sinc function of the square of a momentum coordinate, and is not in most dictionaries of transforms.

\section{The Double-Gaussian Approximation}
In what follows here, we approximate the Sinc-Gaussian biphoton momentum-space wavefunction \eqref{SGmomWavf} as a Double-Gaussian function (as seen in \cite{LawEberly2004} and\cite{Fedorov2009}), by matching the second order moments in the sums and differences of the transverse momenta. Transforming this approximate wavefunction to position space, and computing the correlation width gives us an estimate of the true correlation width seen experimentally that we later compare with more exact calculations and experimental data. In addition, we take a moment to explore the conveniences that come with the Double-Gaussian wavefunction.

In this analysis, we consider only the horizontal components of the transverse momenta, since the statistics are identical (with our approximations) in both transverse dimensions. The transverse pump profile is already assumed to be a Gaussian. Our first step is to transform to a rotated set of coordinates to separate the Sinc function from the Gaussian.

Let
\begin{equation}
k_{+}=\frac{k_{1x}+k_{2x}}{\sqrt{2}}\qquad\text{and}\qquad k_{-}=\frac{k_{1x}-k_{2x}}{\sqrt{2}}.
\end{equation}
With these rotated coordinates, the (horizontal) biphoton wavefunction becomes:
\begin{equation}
\phi(k_{+},k_{-})= \mathcal{N}\text{sinc}\bigg(\frac{L_{z}\lambda_{p}}{4\pi n_{p}}k_{-}^{2}\bigg)e^{-2\sigma_{p}^{2}k_{+}^{2}}.
\end{equation}

Taking the modulus-squared and integrating over $k_{+}$, we isolate the probability density for $k_{-}$:
\begin{equation}\label{sincDensity}
\rho(k_{-})= \frac{3}{4}\sqrt{\frac{a}{\pi}}\text{sinc}^{2}(a k_{-}^{2})
\end{equation}
where $a\equiv\frac{L_{z}\lambda_{p}}{4\pi n_{p}}$, for convenience. $\rho(k_{-})$ is an even function, so its first-order moment, the expectation $\langle k_{-}\rangle=0$. The second-order moment is nonvanishing, with a value $\langle k_{-}^{2}\rangle=\frac{3}{4a}$. With this second order moment, we can fit $\rho(k_{-})$ to a Gaussian by matching these moments. In doing so, $\rho(k_{-})$ is approximately a Gaussian with width $\sigma_{k_{-}}\equiv \sqrt{\frac{3}{4a}}$. 

To see how good this Gaussian approximation of $\rho(k_{-})$ is, we show in Fig.~2, both the Sinc-based probability density (in momentum space) and the approximate Gaussian density with matched moments. The overall scale of the central peak is captured but the shape is significantly different. However, a Gaussian probability density for the position difference density, $\rho(x_{-})$, appropriately scaled is a good approximation for values near the central peak (though the oscillatory behavior of the wings is still not captured). In Fig.~3, we plot various choices of an approximate Gaussian density for our transformed Sinc-based position difference density function $\rho(x_{-})$. We find that simply setting the central maxima of both densities equal to each other works very well, as discussed in the next section.

\begin{figure}[t]
\centerline{\includegraphics[width=\columnwidth]{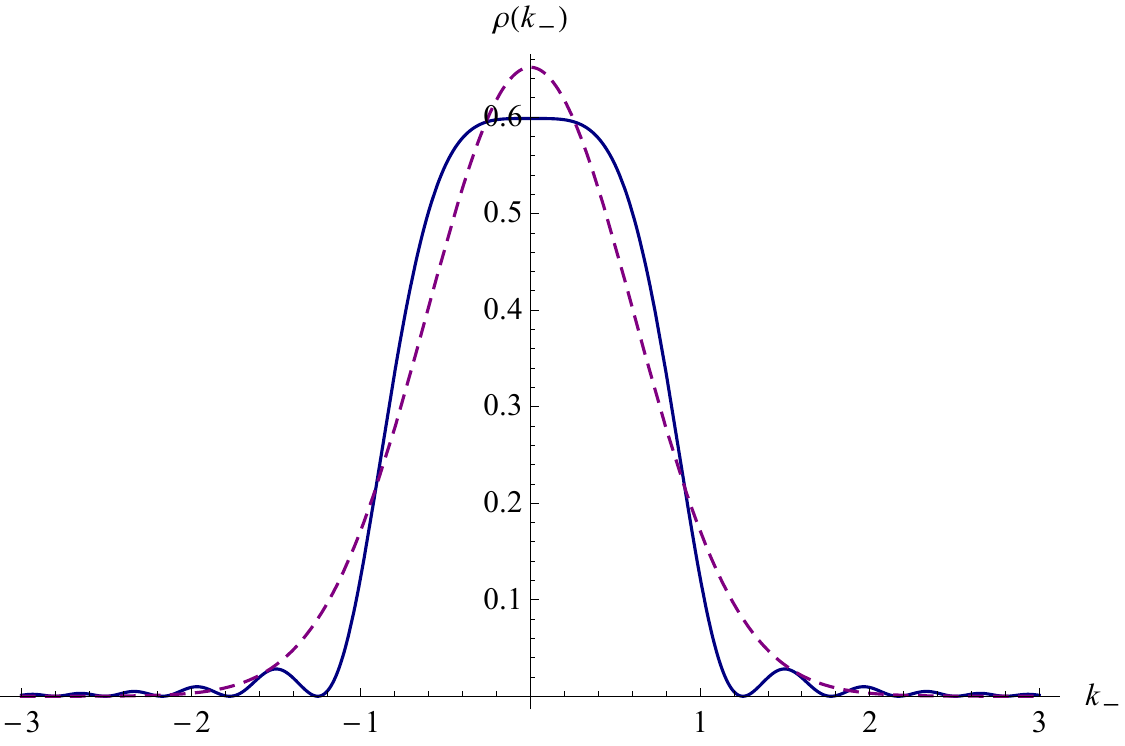}}
\caption{Plot comparing estimates of the momentum difference probability density $\rho(k_{-})$. The solid (blue) curve with wavy side-bands gives our Sinc-based probability density estimate \eqref{sincDensity}, where we set $a=2$ for convenience. The dashed (magenta) Gaussian  curve gives our Gaussian-based probability density estimate with matching means and variances.}
\end{figure}

By approximating the Sinc-Gaussian wavefunction as a Double-Gaussian wavefunction, the inverse Fourier transform to position space becomes very straightforward. We note that $k_{+}$ and $k_{-}$ form an orthogonal pair of coordinates, as $k_{1x}$ and $k_{2x}$ do. Because of this, the inverse Fourier transform is separable \footnote{The Fourier transform convention we use is the unitary convention: $\tilde{\psi}(k_{1x},k_{2x}) = \frac{1}{2\pi}\iint dx_{1}dx_{2}\;e^{-i (x_{1}k_{1x}+x_{2}k_{2x})}\psi(x_{1},x_{2})$, and $\psi(x_{1},x_{2}) = \frac{1}{2\pi}\iint dk_{1x}dk_{2x}\;e^{i (x_{1}k_{1x}+x_{2}k_{2x})}\tilde{\psi}(k_{1x},k_{2x})$. Since the Fourier transform is invariant under rotations (i.e., since the argument in the exponential can be thought of as an inner product between two vectors), we get identical formulas for the Fourier transform in rotated coordinates. In particular, we find $\psi(x_{+},x_{-}) = \frac{1}{2\pi}\iint dk_{+}dk_{-}\;e^{i (x_{+}k_{+}+x_{-}k_{-})}\tilde{\psi}(k_{+},k_{-})$.}, and we find:
\begin{equation}\label{psixdgauss}
\psi(x_{+},x_{-})\approx \bigg(\frac{1}{\sqrt{2\pi \sigma_{x_{+}} \sigma_{x_{-}}}}\bigg)e^{-\frac{x_{-}^{2}}{4\sigma_{x_{-}}^{2}}}e^{-\frac{x_{+}^{2}}{4\sigma_{x_{+}}^{2}}}.
\end{equation}
where $\sigma_{x_{-}}^{2}=\frac{1}{4\sigma_{k_{-}}^{2}}$, $\sigma_{x_{+}}^{2}=\frac{1}{4\sigma_{k_{+}}^{2}}=2\sigma_{p}^{2}$, and where
\begin{equation}
x_{+}=\frac{x_{1}+x_{2}}{\sqrt{2}}\qquad\text{and}\qquad x_{-}=\frac{x_{1}-x_{2}}{\sqrt{2}}.
\end{equation}

\begin{figure}[t]
\centerline{\includegraphics[width=\columnwidth]{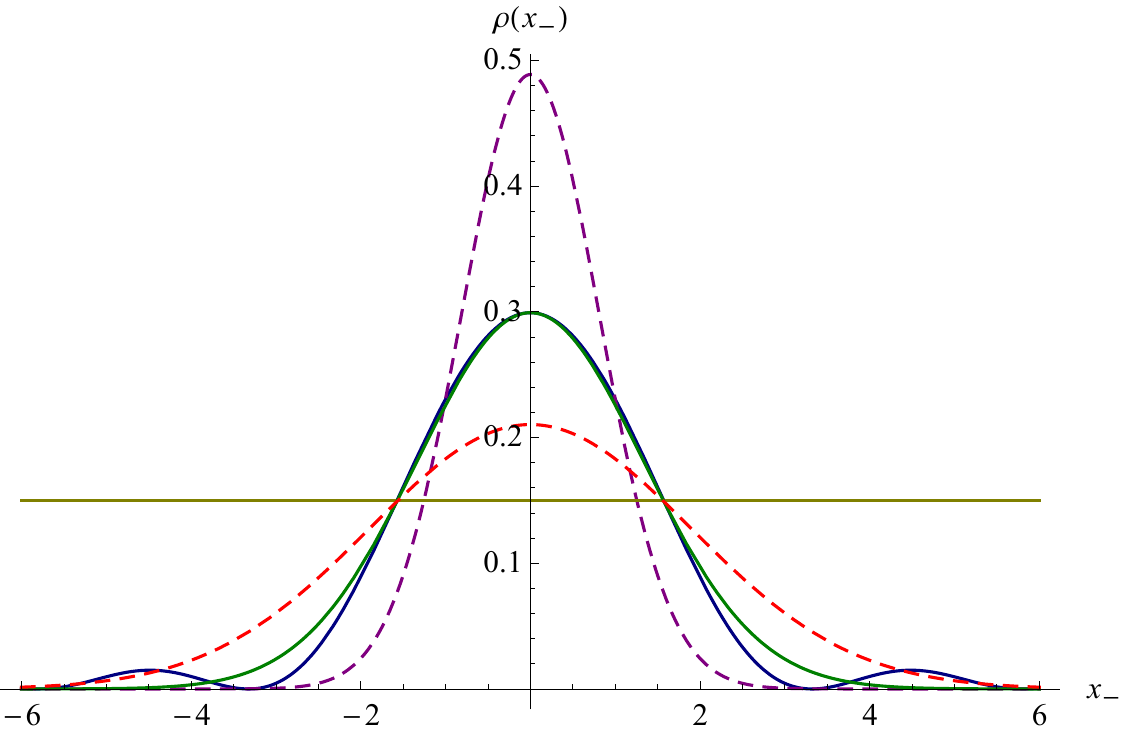}}
\caption{Plot comparing different estimates of $\rho(x_{-})$. The solid blue wavy curve is our most accurate estimate from the transformed Sinc-based distribution \eqref{FinkFunction}. The tall dashed (magenta) curve is the Gaussian distribution obtained from matching momentum means and variances, while the shallow dashed (red) curve is the Gaussian distribution obtained by matching position means and variances. The solid (green) curve gives us a refined Gaussian approximation, by setting the central maximums equal to one another. The flat (gold) line, gives the height of the half maximum of the Sinc-based probability density \eqref{FinkFunction} (blue curve). We see that the widths of half maximum are nearly identical (off by less than $0.3\%$) for the Sinc-based and refined Gaussian distributions Again, we set $a=2$ for convenience.}
\end{figure}

\subsection{Usefulness of the Double-Gaussian approximation}
Here, we digress to discuss the usefulness of the Double-Gaussian approximation. To begin, we express $x_{+}$ and $x_{-}$ in terms of $x_{1}$ and $x_{2}$, and take the magnitude-squared of $\psi(x_{1},x_{2})$ to get a Double-Gaussian probability density $\rho^{DG}$:
\begin{equation}\label{xdgauss}
\rho^{DG}(x_{1},x_{2})=\bigg(\frac{1}{2\pi \sigma_{x_{+}} \sigma_{x_{-}}}\bigg) e^{-\frac{(x_{1}-x_{2})^{2}}{4\sigma_{x_{-}}^{2}}}e^{-\frac{(x_{1}+x_{2})^{2}}{4\sigma_{x_{+}}^{2}}}.
\end{equation}

\begin{figure}[t]
\centerline{\includegraphics[width=\columnwidth]{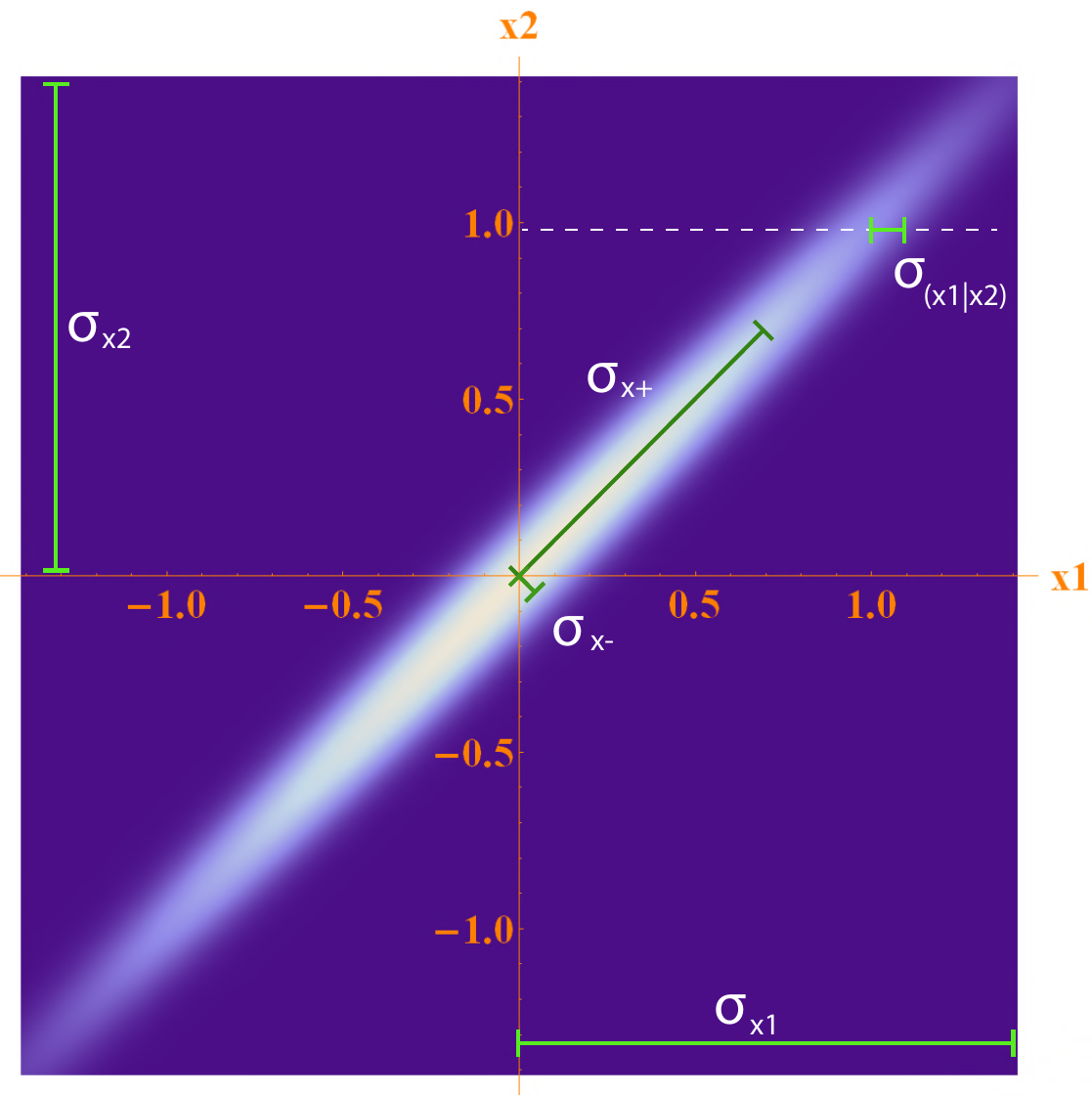}}
\caption{Plot of the Double-Gaussian probability density for $\sigma_{x_{+}}=1$ unit and $\sigma_{x_{-}}=0.075$ units. The horizontal dotted line indicates a particular value of $x_{2}$, so that the half-width of the Gaussian along that dotted line is the conditioned half-width $\sigma_{(x_{1}|x_{2})}$. The equivalent Sinc-Gaussian distribution will have subtle side bands  (at about $5.0\%$ the maximum intensity) parallel to the long axis of this Double-Gaussian as in Fig.~3.}
\end{figure} 

\begin{table}
\centering
\caption{Statistics of the Double Gaussian}
\begin{tabular}{l|p{5cm}}
\hline \hline Name  & Value \\
\hline	
marginal means   & $\langle x_{1}\rangle=\langle x_{2}\rangle=0$  \\[1ex]
\cline{1-1}
conditioned mean   & $\langle x_{2}\rangle_{\rho(x_{2}|x_{1})}= x_{1}-\frac{2 x_{1} \sigma_{x_{-}}^{2}}{\sigma_{x_{+}}^{2}+\sigma_{x_{-}}^{2}}=r x_{1}$ \\[1ex]
\cline{1-1}
marginal variance  & $\sigma_{x_{1}}^{2}=\sigma_{x_{2}}^{2}=\frac{\sigma_{x_{+}}^{2}+\sigma_{x_{-}}^{2}}{2}$  \\[1ex]
\cline{1-1}
conditioned variance & $\sigma_{(x_{1}|x_{2})}^{2}=\sigma_{(x_{2}|x_{1})}^{2}=\frac{2\sigma_{x_{+}}^{2}\sigma_{x_{-}}^{2}}{\sigma_{x_{+}}^{2}+\sigma_{x_{-}}^{2}}$  \\[1ex]
\cline{1-1}
co-variance & $\langle x_{1}x_{2}\rangle-\langle x_{1}\rangle\langle x_{2}\rangle=\frac{\sigma_{x_{+}}^{2}-\sigma_{x_{-}}^{2}}{2}$  \\[1ex]
\cline{1-1}
Pearson r value & $r=\frac{\langle x_{1}x_{2}\rangle-\langle x_{1}\rangle\langle x_{2}\rangle}{\sigma_{x_{1}}\sigma_{x_{2}}}=\frac{\sigma_{x_{+}}^{2}-\sigma_{x_{-}}^{2}}{\sigma_{x_{+}}^{2}+\sigma_{x_{-}}^{2}}$  \\[1ex]
\cline{1-1}
joint entropy & $h(x_{1},x_{2})=\log(2\pi e\sigma_{x_{+}}\sigma_{x_{-}})$  \\[1ex]
\cline{1-1}
marginal entropy   & $h(x_{1})=\frac{1}{2}\log(\pi e(\sigma_{x_{+}}^{2}+\sigma_{x_{-}}^{2}))$;\\ & $\qquad\;\;=h(x_{2})$ \\[1ex]
\cline{1-1}
mutual information & $h(x_{1}\!\!:\!\!x_{2})=h(x_{1})+h(x_{2})-h(x_{1},x_{2})$;\\&
$=\log\bigg(\frac{\sigma_{x_{+}}^{2}+\sigma_{x_{-}}^{2}}{2\sigma_{x_{+}}\sigma_{x_{-}}}\bigg)$
$=\log\big(\frac{\sigma_{x_{1}}}{\sigma_{(x_{1}|x_{2})}}\big)$ \\[1ex] 
\hline \hline
Probability notation & $\sigma^{2}_{(x_{1}|x_{2})}\equiv\sigma^{2}_{\rho(x_{1}|x_{2})}$\\
 & $\rho(x_{1}|x_{2})=\frac{\rho(x_{1},x_{2})}{\rho(x_{2})}$\\
\hline\hline
\end{tabular}
\label{table1}
\end{table}

Here, we define the transverse correlation width as the standard deviation of the distance between $x_{1}$ and $x_{2}$ (i.e., $\sigma_{(x_{1}-x_{2})}$). This is not defined as a half-width, since it represents the full width (at $\frac{1}{\sqrt{e}}$ of the maximum) of the signal and idler photons' position distributions conditioned on the location of the prior pump photon (see Section VI). For the Double-Gaussian density, the transverse correlation width, $\sigma_{(x_{1}-x_{2})}$, is $\sqrt{2}\sigma_{x_{-}}$.

Alternatively, the Double-Gaussian density can be put into the standard form of a bi-variate Gaussian density function;
\begin{equation}
\rho^{DG}(x_{1},x_{2})= \bigg(\frac{\sqrt{ac-b^{2}}}{\pi}\bigg) e^{- (a x_{1}^{2}+2b x_{1}x_{2}+c x_{2}^{2})},
\end{equation}
where
\begin{align}
a&=c=\frac{\sigma_{x_{+}}^{2}+\sigma_{x_{-}}^{2}}{4\sigma_{x_{+}}^{2}\sigma_{x_{-}}^{2}},\\
b&=\frac{\sigma_{x_{-}}^{2}-\sigma_{x_{+}}^{2}}{4\sigma_{x_{+}}^{2}\sigma_{x_{-}}^{2}}.
\end{align}
This Double-Gaussian probability density has a number of useful properties. First, it is separable into single Gaussians in rotated coordinates, making many integrals straightforward to do analytically. Second, the marginal and conditional probability densities of the Double Gaussian density function are also Gaussian density functions. Because of this, many statistics of the Double-Gaussian density have particularly simple forms. For examples, consider the statistics in Table~1.

In addition, the Double-Gaussian is uniquely defined by its marginal and conditioned means and variances. As seen in Fig.~4, these values give a straightforward characterization of the overall shape of the Double-Gaussian distribution. 

\subsubsection{Fourier-Transform Limited properties of the Double-Gaussian}
Gaussian wavefunctions are minimum uncertainty wavefunctions in that they are Fourier-transform limited; Heisenberg's uncertainty relation;
\begin{equation}
\sigma_{x}\sigma_{k}\geq \frac{1}{2},
\end{equation}
is satisfied with equality. The Double-Gaussian wavefunction \eqref{xdgauss} factors into a product of two Gaussians (one in $x_{+}$ and the other in $x_{-}$), and so the standard deviations of these rotated coordinates also saturate the Heisenberg relation:
\begin{equation}
\sigma_{x_{+}}\sigma_{k_{+}}=\frac{1}{2}\qquad:\qquad\sigma_{x_{-}}\sigma_{k_{-}}=\frac{1}{2}.
\end{equation}
Remarkably, the simple expressions for the statistics of the double-Gaussian distribution (see Table \ref{table1}) show that these are not the only pairs that are related this way. Since conditioning measurements on a single ensemble of events $\lambda$ doesn't change the fact that those measurements must satisfy an uncertainty relation, we find:
\begin{equation}
\sigma_{(x_{1}|\lambda)}\sigma_{(k_{1}|\lambda)}\geq\frac{1}{2}
\end{equation}
is still a valid uncertainty relation. In addition, since conditioning on average reduces the variance \footnote{That conditioning on average reduces the variance can be seen from the law of total variance. Given two random variables $X$ and $Y$, the variance of $X$ is equal to the mean over $Y$ of the conditioned variance $Var(X|Y)$ plus the variance over $Y$ of the conditioned mean $E[X|Y]$. Both of these terms are non negative, so the mean conditioned variance never exceeds the unconditioned variance.}, we arrive at the relations:
\begin{equation}\label{CondUncRel}
\sigma_{x_{1}}\sigma_{(k_{1}|k_{2})}\geq\frac{1}{2}\qquad:\qquad \sigma_{k_{1}}\sigma_{(x_{1}|x_{2})}\geq\frac{1}{2}.
\end{equation}
These relations are also useful for understanding how narrowband frequency filters undermine the resolution of temporal correlations as discussed in Appendix C. For the Double-Gaussian state, these relations are saturated as well. From these properties, we may find many other useful identities for the double-Gaussian including:
\begin{align}
r_{x}&=-r_{k}\\
\frac{\sigma_{x_{+}}}{\sigma_{x_{-}}}=\frac{\sigma_{k_{-}}}{\sigma_{k_{+}}}\qquad &: \qquad \frac{\sigma_{x_{1}}}{\sigma_{(x_{1}|x_{2})}}=\frac{\sigma_{(k_{1}|k_{2})}}{\sigma_{k_{1}}}
\end{align}
where $r_{x}$ and $r_{k}$ is the Pearson correlation coefficient for the position and momentum statistics of the Double-Gaussian, respectively.

\subsubsection{Propagating the Double-Gaussian field}
One especially useful aspect of the Double-Gaussian wavefunction, is that it is simple to propagate (in the paraxial regime). Given the transverse momentum amplitude profile of a nearly monochromatic optical field in one transverse plane, we can find the transverse momentum profile at another optical plane by multiplying it by the paraxial free-space transfer function $T_{fs}(z:k_{x},k_{y})$ \footnote{The free space transfer function comes about due to the momentum decomposition of an optical field being a sum (or integral) over plane waves. For each plane wave defined by $k_{x}$, $k_{y}$, and $k_{z}$, we add a phase corresponding to the plane wave translating a total forward distance $z$. The particular form of the free space transfer function used here is due to the small angle- or paraxial approximation. For a good reference on this topic, see \cite{goodman1968introduction}.}:
\begin{equation}
T_{fs}(z:k_{x},k_{y})=e^{ikz -\frac{iz}{2k}(k_{x}^{2}+k_{y}^{2})}.
\end{equation}
For an entangled pair of optical fields at half the pump frequency, the full transfer function becomes:
\begin{align}
&T_{fs}(z_{1},z_{2}:k_{1x},k_{1y},k_{2x},k_{2y})=\nn\\
&=e^{i\frac{k_{p}}{2}(z_{1}+z_{2}) -\frac{iz_{1}}{k_{p}}(k_{1x}^{2}+k_{1y}^{2})}e^{-\frac{iz_{2}}{k_{p}}(k_{2x}^{2}+k_{2y}^{2})}.
\end{align}
Since a global constant phase $e^{i\frac{k_{p}}{2}(z_{1}+z_{2})}$ can come outside the Fourier transform integral, and the relative phases and amplitudes in position space will be independent of this factor, we can remove it from the transfer function, and express the remaining transfer function simply as a product of a horizontal and vertical transfer function:
\begin{align}
&T_{fs}(z_{1},z_{2}:k_{1x},k_{1y},k_{2x},k_{2y})=\nn\\
&=T_{fsx}(z_{1},z_{2}:k_{1x},k_{2x})T_{fsy}(z_{1},z_{2}:k_{1y},k_{2y}).
\end{align}
where
\begin{equation}
T_{fsx}(z_{1},z_{2}:k_{1x},k_{2x})=e^{-\frac{iz_{1}k_{1x}^{2}}{k_{p}}}e^{-\frac{iz_{2}k_{2x}^{2}}{k_{p}}},
\end{equation}
and $T_{fsy}(z_{1},z_{2}:k_{1y},k_{2y})$ is similarly defined.

Because our position-space wavefunction is also (approximately) separable into a product of vertical and horizontal wavefunctions, we can propagate $\psi^{DG}(x_{1},x_{2})$ (i.e., the Double-Gaussian approximation to $\psi(x_{1},x_{2})$), without having to first propagate the entire transverse wavefunction. Doing so, gives us:
\begin{equation}
\psi^{DG}(x_{1},x_{2}:z_{1},z_{2})=\mathcal{F}^{-1}[\tilde{\psi}^{DG}(k_{1x},k_{2x})e^{-\frac{iz_{1}k_{1x}^{2}}{k_{p}}}e^{-\frac{iz_{2}k_{2x}^{2}}{k_{p}}}],
\end{equation}
where $\mathcal{F}^{-1}$ is the inverse-Fourier transform operator. Perhaps surprisingly, propagating the double-Gaussian field simply gives another bi-variate Gaussian density (see in appendix, \eqref{DGPropagated2varb}). What changes through propagation is the parameters defining the Double-Gaussian field. Taking \eqref{psixdgauss} to be the biphoton field backpropagated to the center of the crystal (where $z_{1}=z_{2}=0$, and $\sigma_{x_{+}}$ and $\sigma_{x_{-}}$ to be parameters defining the field at $z_{1}=z_{2}=0$, the transverse probability density of the photon pair when the propagation distances are equal (i.e., $z_{1}=z_{2}=z$), as they would be if we measure both fields is the same image plane, gives us:
\begin{align}\label{rhoDGsimpZ}
\rho^{DG}(x_{1},x_{2};z)&= \bigg(\frac{1}{2\pi \tilde{\sigma}_{x_{+}}(z) \tilde{\sigma}_{x_{-}}(z)}\bigg) e^{-\frac{(x_{1}-x_{2})^{2}}{4\tilde{\sigma}_{x_{-}}(z)^{2}}}e^{-\frac{(x_{1}+x_{2})^{2}}{4\tilde{\sigma}_{x_{+}}(z)^{2}}}\\
:\tilde{\sigma}_{x_{+}}(z)&\equiv\sqrt{\sigma_{x_{+}}^{2}+\bigg(\frac{z}{\sigma_{x_{+}}k_{p}}\bigg)^{2}},\\
:\tilde{\sigma}_{x_{-}}(z)&\equiv\sqrt{\sigma_{x_{-}}^{2}+\bigg(\frac{z}{\sigma_{x_{-}}k_{p}}\bigg)^{2}}.
\end{align}
This particularly illustrates the convenience of working with the Double-Gaussian density, as we need only find effective values for $\sigma_{x_{+}}$ and $\sigma_{x_{-}}$ at one distance $z$ to see how the biphoton field might change under propagation to the same imaging plane \footnote{On the other hand, propagating to independent imaging planes $z_{1}$ and $z_{2}$ is a more elaborate result discussed in the Appendix.}. In particular, where $\sigma_{x_{+}}$ is much larger than $\sigma_{x_{-}}$ for highly entangled light, we can see that the position correlations (say, as measured by the Pearson correlation coefficient) decrease to zero, and gradually become strong anti-correlations as we move to the far field. This does not imply, however, that the photon pairs dis-entangle and re-entangle under propagation \cite{eberly_TransverseMig_PRA2007}; the entanglement migrates to the relative phase of the joint wavefunction and back again. Throughout the rest of this paper, all transverse correlation widths, probability densities, and biphoton amplitudes will be assumed to be taken at $z_{1}=z_{2}=0$ (or an image plane conjugate to this plane), unless otherwise specified.

\section{Estimating the Transverse Correlation Width}
Using our earlier notation, the approximation to the Double-Gaussian wavefunction is expressed as follows:
\begin{equation}\label{phiDGauss}
\psi(x_{1},x_{2})\approx \sqrt\frac{1}{2\sqrt{2}\pi \sigma_{p} \sigma_{x_{-}}} e^{-\frac{(x_{1}-x_{2})^{2}}{8\sigma_{x_{-}}^{2}}}e^{-\frac{(x_{1}+x_{2})^{2}}{16\sigma_{p}^{2}}},
\end{equation}
where $\sigma_{x_{-}}=\sqrt{\frac{a}{3}}$, making the transverse correlation width, $\sigma_{(x_{1}-x_{2})}=\sqrt{\frac{2a}{3}}$.

To see just how good (or not) this Gaussian-based estimate of $\sigma_{(x_{1}-x_{2})}$ is, we compare our Gaussian approximation to the the probability density of $x_{-}$ obtained when taking the Fourier transform of the Sinc-based function of $k_{-}$ \eqref{sincDensity}. The more accurate probability density obtained from that Fourier transform is:
\begin{align}\label{FinkFunction}
\rho(x_{-})&=\frac{3}{16\sqrt{\pi a^{3}}}\bigg|x_{-}\sqrt{2\pi}\bigg(\mathcal{S}\bigg(\frac{x_{-}}{\sqrt{2\pi a}}\bigg)-\mathcal{C}\bigg(\frac{x_{-}}{\sqrt{2\pi a}}\bigg)\bigg)+\nn\\
&+ 2\sqrt{a} \bigg(\cos\bigg(\frac{x_{-}^{2}}{4a}\bigg)+\sin\bigg(\frac{x_{-}^{2}}{4a}\bigg)\bigg)\bigg|^{2},
\end{align}
where $\mathcal{C}(x)$ and $\mathcal{S}(x)$ are the Fresnel integrals, integrating over $\cos(\frac{\pi}{2}t^{2})$ and $\sin(\frac{\pi}{2}t^{2})$, respectively. As seen in Fig.3, the Gaussian approximation obtained by matching $\langle k_{-}\rangle$ and $\langle k_{-}^{2}\rangle$ gives a full width at half maximum (FWHM) within an order of magnitude of the FWHM of the more accurate approximation \eqref{FinkFunction}.

However, with a width $\sigma_{x_{-}}$ of $\sqrt{\frac{8a}{9}}$ (i.e., by setting the maximums of our (Sinc-based and Gaussian-based) approximate density functions equal to one another), where again, $a\equiv\frac{L_{z}\lambda_{p}}{4\pi n_{p}}$, one obtains a FWHM only $0.3\%$ smaller than the FWHM from the more accurate case \eqref{FinkFunction}. Indeed, numerical estimates based on fitting the widths of the Sinc-based density to the Gaussian density  have been performed \cite{LawEberly2004,eberly_TransverseMig_PRA2007} to great effect \cite{edgar2012imaging}. Since choosing which width to fit is somewhat arbitrary, we point out that the peak-matching fit also fits the full width at $48.2\%$ of the maximum. However, the best estimate of $\sigma_{x_{-}}$ is obtained by an explicit calculation of $\sigma_{x_{-}}\equiv \sqrt{\langle x_{-}^{2}\rangle}$ from the more accurate density \eqref{FinkFunction}. Remarkably, we find that $\sigma_{x_{-}}$ is simply $\sqrt{\frac{9a}{5}}$, which in turn gives us a transverse correlation width, $\sigma_{(x_{1}-x_{2})}$, of $\sqrt{\frac{18a}{5}}$. In addition, matching this exact variance to define an approximate Double-Gaussian wavefunction also gives us the maximum likelihood estimate (i.e., the estimated distribution with minimum relative entropy to the more accurate model) of a Double Gaussian distribution fitting our more exact results. As a summary of our calculations, see the following:
\begin{equation}\label{widthest}
\boxed{
\begin{array}{rl}
\sigma^{(\text{exact})}_{(x_{1}-x_{2})}&=\sqrt{2}\sigma^{(\text{exact})}_{x_{-}}=\sqrt{\frac{18a}{5}}=\sqrt{\frac{9 L_{z}\lambda_{p}}{10 \pi n_{p}}},\\
\sigma^{(PM)}_{(x_{1}-x_{2})}&=\sqrt{2}\sigma^{(PM)}_{x_{-}}=\sqrt{\frac{16a}{9}}=\sqrt{\frac{4 L_{z}\lambda_{p}}{9 \pi n_{p}}}.
\end{array}}
\end{equation}
Here, $\sigma^{PM}$ refers to the peak-matching estimate that also nearly matches the widths of the Gaussians, while $\sigma^{(\text{exact})}$ is our more accurate calculation. Both estimates have their uses when examining experimental data, as we shall see.

Though explicitly calculating the variance of $x_{-}$ according to our accurate density function \eqref{FinkFunction} gives us the best possible estimate of $\sigma_{x_{-}}$, it does not necessarily give us the best fitting Gaussian approximation to the Sinc-based distribution. The Gaussian obtained by explicitly matching position means and variances, gives a distribution about $42.3\%$ wider than the close fitting distribution we obtain by matching peak values (see Fig. 3 for comparison). The resulting (overly wide) scaled Gaussian distribution  (by matching variances) does not accurately reflect the probabilities of the most likely outcomes (near $x_{-}=0$) (e.g., that within $\pm$ one "sigma", we should get approximately $68\%$ of the total data). Indeed, by setting the central maximums equal to one another, we also find the Gaussian cumulative distribution function (CDF) that most accurately resembles the CDF of our more accurate distribution \eqref{FinkFunction} near its median. As an example of the accuracy of this approximation, our peak-matching approximate Gaussian distribution, gives a total probability within one $\sigma_{x_{-}}$ from the origin of $68.3\%$, while the more accurate density function gives a probability of $69.0\%$, (an absolute difference of only $0.7\%$) over the same interval.

\subsection{Comparison with experimental data}
Though our estimate of $\sigma_{(x_{1}-x_{2})}$ follows from reasonable approximations that work well within typical experimental setups, it is important to show just how well (or not) this estimate of the transverse correlation width corresponds with experimental data. This can be done in (at least) two ways. First, in \cite{Howell2004}, they found a measure of the transverse correlation width by placing a 40$\mu$m slit in the signal beam, and scanning over the idler beam with another 40$\mu$m slit (see Fig.~5 for a diagram of the idealized setup). By measuring coincident detections as they scan, and normalizing the resulting histogram, they measured the conditional transverse probability distribution, and obtained a conditional width, which is approximately identical to the correlation width \footnote{When $\rho(x_{1},x_{2})=\rho(x_{+})\rho(x_{-})\text{ and }\sigma_{x_{+}}\gg\sigma_{x_{-}}$, it follows that $\sigma_{(x_{1}-x_{2})}\approx \sigma_{(x_{1}|x_{2})}$.}. With their measurements, they obtained a transverse correlation width (adjusting for our conventions) of about 13.5$\mu$m (with an estimated error larger than $10\%$), so that our theoretical estimate \eqref{widthest} of 11.6$\mu$m (using their pump wavelength of 390nm and crystal thickness of 2mm) underestimates this by $14.1\%$.

\begin{figure}[t]
\centerline{\includegraphics[width=\columnwidth]{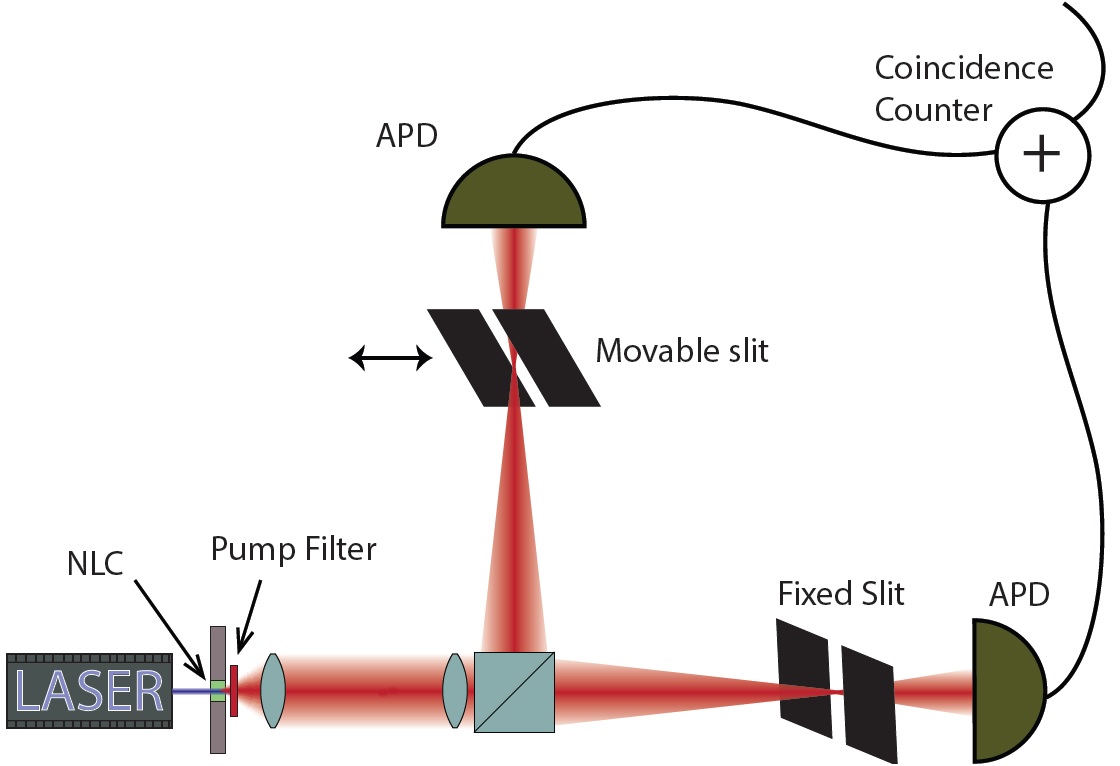}}
\caption{Idealized diagram of an experiment \cite{Howell2004} to measure the transverse correlation width. The nonlinear crystal (NLC) is just after the laser, with a pump filter placed just after that. The beam is broken into signal and idler with a $50/50$ beamsplitter. There is a loss of coincidences due to the beamsplitter but this doesn't affect the spatial intensity profile. Two lenses are used to image the exiting face of the nonlinear crystal onto the image planes where two slits are placed. With one slit fixed, the other mobile, and Avalanche Photo-Diodes (APDs) behind each slit, one can obtain the correlation width by comparing the width of the coincidence distribution to the width of the slits.}
\end{figure}

Another measurement with the same laser and crystal was taken in \cite{bennink_PhysRevLett.92.033601}, where they obtained a transverse correlation width of $17\pm 7\mu m$. Given how these experiments' resolutions were limited both by finite slit widths \footnote{In order to obtain an estimate of 13.5$\mu$m for the transverse correlation width using slits 40$\mu$m wide, they deconvolved their coincidence histograms with the slit rectangle function. This gave them more accurate estimates for the joint coincidence distributions at arbitrary resolution from which they could obtain more accurate estimates of the transverse correlation width.} and a large statistical uncertainty in $\sigma_{(x_{1}-x_{2})}$, our approximation is accurate to within experimental uncertainty. More recently, \cite{edgar2012imaging}, an experiment was performed in which the joint position photon distribution was imaged with a camera. By fitting a Double-Gaussian to their empirical distribution, they found a correlation width of $10.9\pm0.7\mu$m (for their 355nm pump beam and 5mm crystal). Surprisingly, this agrees more with our peak-matching estimate of $12.2\mu$m than with our ostensibly more accurate estimate of $17.5\mu$m (for these parameters). This however is to be expected, as the fitting by its very nature gives a result whose shape most closely resembles the shape the data gives, and low-level noise will mask information about the distribution beyond the central peak. Future experiments with higher-resolution measurements are needed to better explore the strength of this approximation. 

The second way that one can use experimental data to place a limit on the transverse correlation width is to use the comparatively larger amount of data about temporal correlation widths. As an example, if one knew that in a single downconversion event, the photons were generated no further than 100fs apart 90$\%$ of the time, then the speed of light assures us that the photon pair could be no farther than 30$\mu$m apart 90$\%$ of the time as well. Indeed, in \cite{AliKhan_ETent_PRA2006}, they measured approximately a 50fs time-correlation width (using our convention) with downconverted photons from the same 390nm pump laser incident on a 2mm long nonlinear crystal. With this value, we can place an upper bound to the transverse correlation width of the light in that setup by 15$\mu$m, which is not substantially above our 14.9$\mu$m estimate.

\section{The Biphoton birth zone}
When a photon pair is created in SPDC, the location of the pair production can be essentially anywhere in the crystal illuminated by the pump beam. The uncertainty in where this pair-production takes place is limited by the uncertainty in the location of the pump photon. However, for any given photon pair created, the mean separation between the two is generally much smaller than the uncertainty (i.e., standard deviation) in the expected location of the downconversion event. It is useful to conceptualize the region surrounding that mean position where the daughter photons are most likely to be found as what we shall call a birth zone.

Given that momentum is very nearly conserved in SPDC, we define the expected (transverse) location of the downconversion event as the mean of the two photons' positions $x_{m}$:
\begin{equation}
x_{m}\equiv \frac{x_{1}+x_{2}}{2}=\frac{x_{+}}{\sqrt{2}}.
\end{equation}
With the Double-Gaussian wavefunction \eqref{phiDGauss} as our model for transverse position statistics in SPDC, we find the standard deviation in the mean position $\sigma_{x_{m}}$ to be:
\begin{equation}
\sigma_{x_{m}}=\frac{1}{2}\sigma_{(x_{1}+x_{2})}=\sigma_{p}.
\end{equation}
\begin{figure}[t]
\centerline{\includegraphics[width=\columnwidth]{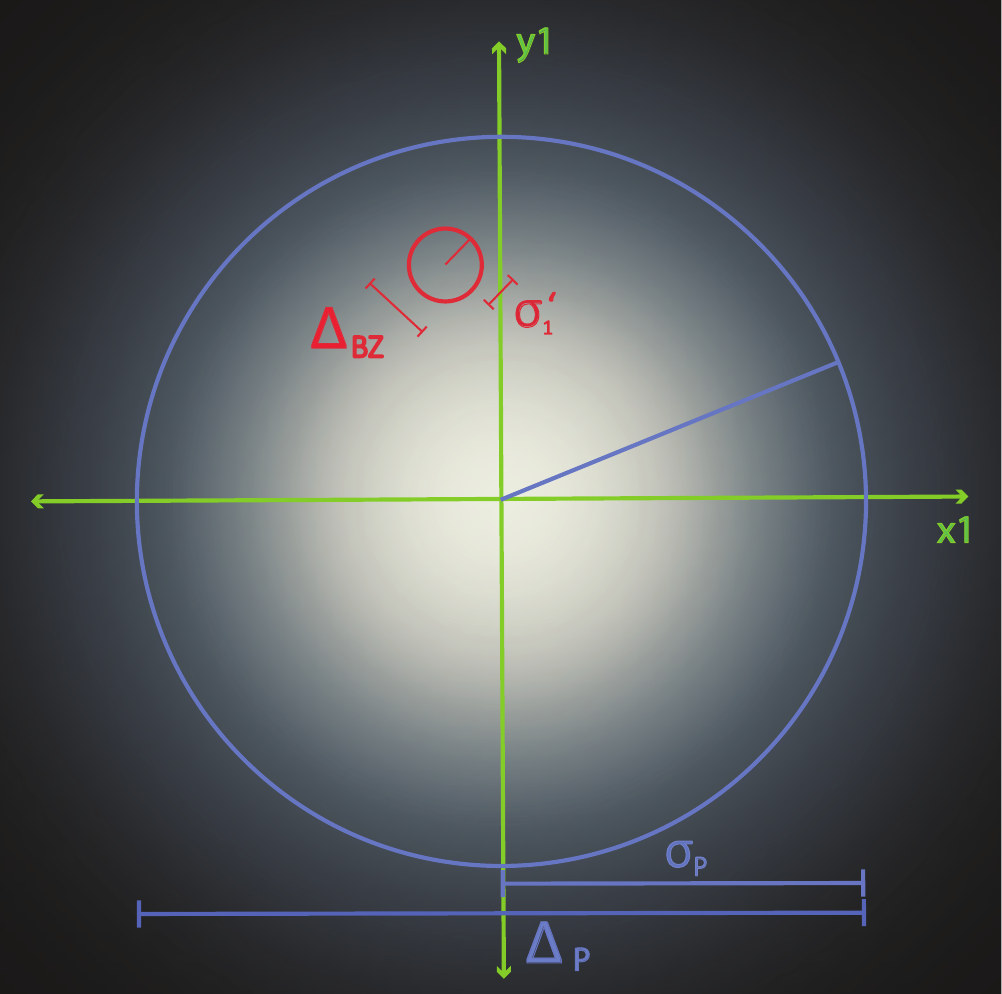}}
\caption{Idealized diagram of the transverse intensity profile (in both $x$ and $y$) of the downconverted light just as it exits the crystal. The blue circle encapsulates the region within one standard deviation of the pump photon position (or also approximately the signal or idler photon position) from the beam center. The red circle is centered on a particular downconversion event, and encapsulates the region where the signal and idler photons are likely to be found given that their mean position is known to be at the canter of the circle (i.e., one birth zone). For a sense of scale, we let $\Delta_{p}/\Delta_{BZ}=10$.}
\end{figure} 
In addition, we define the width of the pump, $\Delta_{P}$, to be twice this standard deviation. While the photon pair is expected to be created at $x_{m}$ with uncertainty $\sigma_{p}$, the signal and idler photons, conditioned on their mean position being at some $x_{m}'$, will each have position uncertainties $\sigma_{x_{1}}'=\sigma_{x_{2}}'=\sigma_{x_{-}}/\sqrt{2}$. As such, we define the birth zone width, $\Delta_{BZ}$, to be twice these conditioned position uncertainties \footnote{The equation $\sigma_{x_{1}}'= \sqrt{2}\sigma_{x_{-}}$ holds only when $\rho(x_{+},x_{-})$ is separable into the product $\rho(x_{+})\rho(x_{-})$, as is the case for both the Double-Gaussian and Sinc-Gaussian distributions.}, giving us:
\begin{equation}
\Delta_{BZ}\equiv 2\sigma_{x_{1}}'= \sqrt{2}\sigma_{x_{-}}=\sigma_{(x_{1}-x_{2})}.
\end{equation} 
With these definitions, we take the birth zone number $N$ (which in one dimension is $\Delta_{P}/\Delta_{BZ}$ or alternatively $\sigma_{x_{+}}/\sigma_{x_{-}}$) to be a measure of the degree of correlation between the signal and idler photon fields \footnote{Although the birth zone number $N$ is one dimension is defined as $\Delta_{P}/\Delta_{BZ}$, we can express it simply in terms of the measured full-width at half-maximum of the pump. Using the peak-matching approximation for the birth zone width, we find: $N=\frac{FWHM_{p}}{\sqrt{\frac{8 \text{ ln}(2)}{9\pi n_{p}}L_{z}\lambda_{p}}}$.}. The birth zone number $N$ is not to be confused with the Fedorov ratio $R$ \cite{FedorovRatioPRA2004}, which is the ratio $\sigma_{x_{1}}/\sigma_{(x_{1}|x_{2})}$. In a sense, each birth zone can be thought of as an independent source of photon pairs. In this way, we can develop an intuitive understanding of the coherence properties of the two-photon fields in SPDC \footnote{For an extensive reference on the relationship between first and second-order coherence in biphoton fields in SPDC, see \cite{Saleh_PRA2000}.}.

In particular, the birth zone width is useful in many calculations involving entangled photon pairs in SPDC. Consider the mutual information of the Double-Gaussian distribution \eqref{xdgauss};
\begin{equation}
h(x_{1}:x_{2})=\log\bigg(\frac{\sigma_{x_{+}}^{2}+\sigma_{x_{-}}^{2}}{2\sigma_{x_{+}}\sigma_{x_{-}}}\bigg)=\log\bigg(\frac{1}{2}\big(N+\frac{1}{N}\big)\bigg).
\end{equation}

Also for the double-Gaussian state, the birth zone number $N$ is related to the Schmidt number $K$, a measure of entanglement \footnote{Given a pure continuous-variable density operator $\hat{\rho}_{AB}=|\psi_{AB}\rangle\langle\psi_{AB}|$ describing the joint state shared by parties $A$ and $B$, the Schmidt number $K$ is the reciprocal of the trace of the square of either reduced density operator; i.e., $K\equiv 1/\text{Tr}[\hat{\rho}_{A}^{2}]=1/\text{Tr}[\hat{\rho}_{B}^{2}]$. The Schmidt number is also known as the inverse participation ratio.} for pure continuous-variable states \cite{LawEberly2004} (see appendix for more details). For this state, $K=\frac{1}{2}(N+1/N)$, and the mutual information becomes:
\begin{equation}\label{MutInfBZnum}
h(x_{1}:x_{2})=\log(K)\approx\log(N)-1.
\end{equation}
where the approximation applies for large $N$.

In addition, the birth zone number gives us a perspective in understanding the tradeoff between the first-order spatial coherence and the measurable biphoton correlations in the downconverted fields \footnote{For a more thorough look into the first-order coherence properties of downconverted fields, see \cite{DixonPRA2010}.}. For completeness, we also briefly discuss the second order spatial coherence, as it is related to the biphoton correlations. As a bit of background, the first order coherence function, $g^{(1)}(a,b)$, is a normalized correlation between the electric field at one point $(a)$ and the electric field at another point $(b)$ (in space or in time). If the electric field is coherent between two points (so that the phase difference between these two points is on average well-defined), then the coherence function will have a magnitude near unity. The second-order coherence function, $g^{(2)}(a,b)$, is a normalized correlation between the intensity (i.e., square of the electric field) at one point with the intensity at another point. While $g^{(1)}(a,b)$ can be used to characterize the extent of interference effects in the signal/idler beams, $g^{(2)}(a,b)$ can be used to characterize the extent of signal/idler photon correlations.

To examine the first and second-order coherence functions in SPDC for a qualitative understanding, we look at the symmetric first order and second order spatial coherence functions, ($g^{(1)}(x,-x)$ and $g^{(2)}(x,-x)$, respectively), as they have particularly simple forms in the Double-Gaussian approximation. The first-order symmetric spatial coherence is defined as: 
\begin{equation}
g^{(1)}(x,-x)\equiv\frac{\langle\hat{a}^{\dagger}_{1}(x)\hat{a}_{1}(-x)\rangle}{\sqrt{\langle\hat{a}^{\dagger}_{1}(x)\hat{a}_{1}(x)\rangle\langle\hat{a}^{\dagger}_{1}(-x)\hat{a}_{1}(-x)\rangle}}, 
\end{equation}
which in terms of our biphoton wavefunction $\psi(x_{1},x_{2})$, can be expressed as:
\begin{equation}
g^{(1)}(x,-x)=\frac{\int dx_{2}\psi^{\ast}(x,x_{2})\psi(-x,x_{2})}{\sqrt{(\int dx_{2}|\psi(x,x_{2})|^{2})(\int dx_{2}|\psi(-x,x_{2})|^{2})}}, 
\end{equation}
We note that the expectation values taken here are taken with the state of the downconverted field $\ket{\Psi_{SPDC}}$.

Similarly, the second-order symmetric spatial coherence is defined as
\begin{equation}
g^{(2)}(x,-x)\equiv\frac{\langle\hat{a}^{\dagger}_{1}(x)\hat{a}^{\dagger}_{2}(-x)\hat{a}_{2}(-x)\hat{a}_{1}(x)\rangle}{\langle\hat{a}^{\dagger}_{1}(x)\hat{a}_{1}(x)\rangle\langle\hat{a}^{\dagger}_{2}(-x)\hat{a}_{2}(-x)\rangle}, 
\end{equation}
which in terms of $\psi(x_{1},x_{2})$, is expressed as:
\begin{equation}
g^{(2)}(x,-x)=\frac{|\psi(x,-x)|^{2}}{(\int dx_{2}|\psi(x,x_{2})|^{2})(\int dx_{1}|\psi(x_{1},-x)|^{2})}.
\end{equation}

Using the Double-Gaussian approximation of the biphoton wavefunction, these symmetric coherence functions take simple Gaussian forms. For $\psi(x_{1},x_{2})$ as defined in \eqref{phiDGauss}, we find:
\begin{equation}
g^{(1)}(x,-x)=e^{-\frac{x^{2}}{2 \Delta_{p}^{2}}\big(\frac{(N^{2}-1)^{2}}{N^{2}+1}\big)}
\end{equation}
and
\begin{equation}
g^{(2)}(x,-x)=\frac{N^{2}+1}{2N}e^{-\frac{x^{2}}{2 \big(\frac{\Delta_{p}}{2N}\big)^{2}}\big(\frac{N^{2}-1}{N^{2}+1}\big)}.
\end{equation}

To find the range of values of $x$ over which $g^{(1)}$ and $g^{(2)}$ are significant, we define the first-order coherence width, $\Delta g^{(1)}$, to be the value of $x$ where $g^{(1)}$ falls to $1/\sqrt{e}$. Note that this can be considered to be "$\sigma$" in a Gaussian probability density \footnote{Since $\Delta g^{(1)}$ is the "$1\sigma$" half-width of $g^{(1)}(x,-x)$, and $g^{(1)}(x,-x)$ gives the coherence of photons separated a distance $2x$ away from one another, photons separated by less than $\Delta g^{(1)}$ will be approximately coherent.}. In addition, we define the second order coherence width, $\Delta g^{(2)}$, as the value of $x$ where $g^{(2)}$ falls below unity, and the correlations can be treated as coming from a nonclassical source of light \footnote{Having a second order coherence $g^{(2)}$ below unity is a witness that the source of light must be nonclassical (i.e., not a coherent or thermal state, as may generated from a collection of independent atoms \cite{loudon2000quantum}).}. With these definitions, we find:
\begin{equation}\label{FrstOrdCohWid}
\Delta g^{(1)}=\Delta_{p}\sqrt{\frac{N^{2}+1}{(N^{2}-1)^{2}}}\approx\quad\xrightarrow{\text{for large }N}\quad\frac{\Delta_{p}}{N}=\Delta_{BZ}
\end{equation}
and
\begin{align}
\Delta g^{(2)}&=\frac{\Delta_{p}}{N}\sqrt{\frac{1}{2}\frac{N^{2}+1}{N^{2}-1}\log\bigg(\frac{N^{2}+1}{2N}\bigg)}\nn\\
&\xrightarrow{\text{for large }N}\quad\approx\frac{\Delta_{p}}{N}\sqrt{\frac{1}{2}\log\bigg(\frac{N}{2}\bigg)}.
\end{align}
We note that the errors in these approximations decrease monotonically, so that for $N>12.3$, the error in our approximation to $\Delta g^{(1)}$ falls below $1\%$, and for $N>11.4$, the error in our approximation to $\Delta g^{(2)}$ also falls below $1\%$.

Based on these calculations of the first and second-order coherence widths, we see that for typical sources of downconversion (where $N\sim 100$) the general area over which the downconverted light will exhibit significant first-order coherence is approximately the same as the area of a birth zone $\Delta_{BZ}^{2}$. Because of this, downconverted light beams having a large degree of position or momentum correlation (i.e, a large N) can be considered as a collection of many independent sources of photon pairs, each incoherent with one another (in the first-order sense). The second-order coherence width tells us a slightly different story since it only differs greatly from the first-order coherence width in the limit of zero correlation, or $N=1$. In this limit, we see that when the first-order coherence width is very large (implying the  downconverted light can be described as a single coherent beam), the second order coherence width is necessarily small. However, the reverse is not true, since for large $N$, one can have small first and second order coherence widths. Since a small second order coherence width could imply either a large or small first-order coherence width, the tradeoff between single photon coherence and biphoton correlation is best understood by comparing $\Delta g^{(1)}$ with $N$ or the mutual information $h(x_{1}\!:\!x_{2})$.

The relationship between the first-order spatial coherence of the downconverted fields, and the biphoton correlations as measured by the mutual information is a manner of tradeoff. The mutual information of the position correlations \eqref{MutInfBZnum} increases with $N$, while $\Delta g^{(1)}$ \eqref{FrstOrdCohWid} decreases with $N$. Thus, highly correlated downconverted fields can be treated as incoherent light, while highly coherent (in the first-order sense) downconverted fields can be treated as an uncorrelated source of downconverted light (i.e., as a single beam at the downconverted frequency producing otherwise uncorrelated photon pairs). As one final point on the relationship between first and second-order coherence, there are cases where the visibilities of first-order and second-order interference are very simply related to one another. In particular, it was shown in \cite{Saleh_PRA2000} that for the two-slit experiment with downconverted biphotons, the first-order visibility $V_{1}$ and the second-order visibility $V_{12}$ follow the relation:
\begin{equation}
V_{1}^{2}+V_{12}^{2}\leq 1.
\end{equation}
This can be understood both in terms of a tradeoff between signal-idler entanglement and single-photon coherence, as well as in terms of the monogamy of entanglement between the signal photon, idler photon, and a measurement device. Indeed, this tradeoff has been used successfully to experimentally estimate the Schmidt number (a measure of entanglement) of photon pairs in a beam of down-converted light \cite{PhysRevA.80.022307}.

\section{Conclusion}
The usefulness of spontaneous parametric downconversion (SPDC) as a source of entangled photon pairs is historically self-evident (see footnote 1). In this discussion, we have looked at the fundamental principles governing SPDC in such a way as is often used in continuous-variable quantum information experiments (namely, degenerate collinear Type-I SPDC). We paid particular attention to how one can predict the transverse-correlation width of photon pairs exiting a nonlinear crystal from first principles with accuracy matching current experimental data. Along the way, we digressed to explore how the double-Gaussian wavefunction used to describe SPDC allows a straightforward analysis of its measurement statistics even under free space propagation. In addition, we have developed further the concept of a biphoton birth zone number, and have shown how it manifests itself in the duality between the correlations within one of the downconverted fields, and the correlations between the downconverted fields. It is our hope that this discussion will inspire further interest in the transverse spatial correlations of entangled photon pairs, and in the relationship between intra- and inter-party coherence.

We gratefully acknowledge  helpful discussions with Dr. Gregory A. Howland, as well as the careful editing of Daniel Lum, Sam Knarr, and Justin Winkler. In addition, we are thankful for the support from the National Research Council, DARPA-DSO InPho Grant No. W911NF-10-1-0404, DARPA-DSO Grant No. W31P4Q-12-1-0015, and AFOSR Grant No. FA9550-13-1-0019.

\appendix
\section{The double-Gaussian field propagated to different distances}
The Double-Gaussian field \eqref{psixdgauss}, when propagated to different distances $z_{1}$ and $z_{2}$ has the same double-Gaussian form, though the coefficients are significantly more complex.
\begin{align}\label{DGPropagated2varb}
\rho^{DG}&(x_{1},x_{2};z_{1},z_{2})\approx \bigg(\frac{\sqrt{ac-b^{2}}}{\pi}\bigg) e^{- (a x_{1}^{2}+2b x_{1}x_{2}+c x_{2}^{2})},\\
:a &= \frac{k_{p}^{2}(\sigma_{x_{+}}^{2}+\sigma_{x_{-}}^{2})(z_{2}^{2}+k_{p}^{2}\sigma_{x_{+}}^{2}\sigma_{x_{-}}^{2})}{d},\nn\\
    :b &= \frac{k_{p}^{2}(\sigma_{x_{+}}^{2}-\sigma_{x_{-}}^{2})(z_{1}z_{2}-k_{p}^{2}\sigma_{x_{+}}^{2}\sigma_{x_{-}}^{2})}{d},\nn\\
    :c&=\frac{k_{p}^{2}(\sigma_{x_{+}}^{2}+\sigma_{x_{-}}^{2})(z_{1}^{2}+k_{p}^{2}\sigma_{x_{+}}^{2}\sigma_{x_{-}}^{2})}{d},\nn\\
    :d&=k_{p}^{2}(z_{1}^{2}+z_{2}^{2})(\sigma_{x_{+}}^{2}+\sigma_{x_{-}}^{2})^{2}+2 k_{p}^{2}z_{1}z_{2}(\sigma_{x_{+}}^{2}-\sigma_{x_{-}}^{2})^{2}+\nn\\ 
    &\qquad +4 z_{1}^{2}z_{2}^{2} + 4 k_{p}^{4}\sigma_{x_{+}}^{4}\sigma_{x_{-}}^{4}.
\end{align}
However, we can still find some useful properties. For example, the Pearson correlation coefficient (see Table~1) has the simple expression:
\begin{equation}
r=\frac{-b}{\sqrt{ac}}.
\end{equation}
In more explicit terms, we get
\begin{equation}
r=r_{0}\frac{1-\bar{z}_{1}\bar{z}_{2}}{\sqrt{(\bar{z}_{1}^{2}+1)(\bar{z}_{2}^{2}+1)}}\quad:\quad r_{0}=\frac{\sigma_{x_{+}}^{2}-\sigma_{x_{-}}^{2}}{\sigma_{x_{+}}^{2}+\sigma_{x_{-}}^{2}},
\end{equation}
where $r_{0}$ is the correlation coefficient at $z_{1}=z_{2}=0$, (see Table~1). In addition, the normalized propagation distances $\bar{z}_{1}$ and $\bar{z}_{2}$ are defined such that $\bar{z}_{1}=z_{1}/(k_{p}\sigma_{x_{+}}\sigma_{x_{-}})$, and $\bar{z}_{2}=z_{2}/(k_{p}\sigma_{x_{+}}\sigma_{x_{-}})$. Using this correlation function for the Double-Gaussian, (as mentioned previously), as we move from the near field ($z_{1},z_{2}\approx 0$) to the far field ($z_{1},z_{2}\gg 0$), the initially strong position correlations gradually weaken and eventually become strong position anti-correlations in the far field. In addition, we also see, that when comparing a measurement of the signal photon in the near field to the idler photon in the far field, the correlations approach zero. Since position measurements in the far field can be taken as measurements of momentum (scaled accordingly), we see that the position of one photon is uncorrelated with the momentum of the other (and vise versa).

\section{Schmidt decomposition and quantum entanglement of the Double-Gaussian state}
In order to \emph{measure} the entanglement of a pair of particles, one needs to know the complete density matrix describing the pair of particles. To reckon with continuous-variable states described by continuous wavefunctions (or mixtures thereof), we need to decompose such wavefunctions into an orthogonal basis of eigenfunctions often corresponding to measurably distinct outcomes of some discrete observable. Occasionally, the density matrix in such a decomposition has a finite expression (with zero amplitudes for all other eigenfunctions in that infinite-dimensional basis), and the analysis is straightforward. Other times, the density matrix has non-trivial components over its entire spectrum, and an exact determination is impossible (though approximations may suffice).

However, when the pair of particles can be described by a pure state (i.e., a single joint wavefunction), it is sufficient to know just the eigenvalues of the reduced density matrix of either particle. These eigenvalues manifest themselves in the Schmidt decomposition of an entangled pure state.

As discussed in \cite{Fedorov2009} and \cite{LawEberly2004}, the Double-Gaussian state can be decomposed into Schmidt modes.
\begin{equation}
\psi^{DG}(x_{1},x_{2})=\sum_{n=0}^{\infty}\sqrt{\lambda_{n}}u_{n}(x_{1})u_{n}(x_{2})
\end{equation}
Here, $u_{n}(x)$ is the $n^{\text{th}}$ ``energy" eigenfunction of the quantum harmonic oscillator, and $\lambda_{n}$, in our notation, is:
\begin{equation}
\lambda_{n}= 4\sigma_{x_{+}}\sigma_{x_{-}}\frac{(\sigma_{x_{+}}-\sigma_{x_{-}})^{2n}}{(\sigma_{x_{+}}+\sigma_{x_{-}})^{2n + 2}}=\frac{4N}{(N+1)^{2}}\bigg(\frac{N-1}{N+1}\bigg)^{2n}
\end{equation}
The Schmidt number $K$ is expressed as the reciprocal of the sum of the squares of the Schmidt eigenvalues:
\begin{equation}
K=\frac{1}{\sum_{n=0}^{\infty}\lambda_{n}^{2}}=\frac{1}{2}\bigg(
N+\frac{1}{N}\bigg).
\end{equation}
Interestingly, since the Schmidt eigenvalues of the Double-Gaussian state are geometrically distributed (a maximum entropy distribution for constant $N$), the Double-Gaussian state is the maximally entangled state for a constant birth zone number.

\section{Heisenberg limited temporal correlations in SPDC}
If instead of taking the small-angle approximation in order to get the transverse biphoton wavefunction \eqref{SGmomWavf}, we look directly at the longitudinal component of the wave vector mismatch $\Delta k_{z}$, we can express the biphoton state in terms of the signal and idler photon frequencies $\omega_{1}$ and $\omega_{2}$ \cite{Mikhailova2008, PhysRevA.78.062327}. In doing so, we find that the frequency (and temporal) correlations between the signal/idler photon pairs differs significantly whether they come from Type-I or Type-II SPDC. In Type-II SPDC, the polarizations of the signal-idler photon pairs are orthogonal to each other, and so experience a different index of refraction in birefringent nonlinear crystals. In this case, the biphoton wavefunction (not counting the pump profile) depends to first order on the difference between the signal and idler photon frequencies, resulting in a sinc function of the frequency difference, which translates to a top-hat function of the time difference. In this case, the temporal correlations can be characterized by the (full) width $W_{(t_{1}-t_{2})}$ of the top-hat function, giving us:
\begin{equation}
\text{(Type-II)}\qquad W_{(t_{1}-t_{2})}=\frac{L_{z}|n_{g}^{(1)}-n_{g}^{(2)}|}{c},
\end{equation}
where $n_{g}^{(1)}$ is the group index of the signal photon at its central frequency $\omega_{p}/2$, and $c$ is the speed of light in vacuum. This width amounts to the accumulated time lag between the signal and idler photons due to their experiencing different indices of refraction. Using the research in \cite{PhysRevLett.77.1917}, where they used a $0.5$mm BBO crystal with a pump wavelength of $351.1$nm, we find that $W_{(t_{1}-t_{2})}\approx 125$fs, which agrees within experimental uncertainty with their results.

In Type-I SPDC, the signal-idler photon pairs have parallel polarizations, and so they experience the same index of refraction in the nonlinear crystal. Consequently, the biphoton wavefunction (not counting pump profile) to lowest order depends on the square of the signal-idler frequency difference. As a result, the techniques used to estimate the transverse spatial correlation width $\sigma_{(x_{1}-x_{2})}$ can also be applied to estimating the temporal correlation width $\sigma_{(t_{1}-t_{2})}$, giving us:
\begin{equation}
\text{(Type-I)}\qquad
\begin{array}{rl}
\sigma^{(exact)}_{(t_{1}-t_{2})}&=\sqrt{\frac{9 L_{z}\kappa_{1}}{10}},\\
\sigma^{(PM)}_{(t_{1}-t_{2})}&=\sqrt{\frac{4 L_{z}\kappa_{1}}{9}}.
\end{array}
\end{equation}
where $\kappa_{1}=\frac{d^{2}k_{1}}{d \omega^{2}}|_{\frac{\omega_{p}}{2}}$ is the group velocity dispersion constant at half the pump frequency.

Because the signal and idler photons experience the same index of refection in Type-I SPDC, their temporal correlation width can be much smaller than for pairs originating from a Type-II source. Indeed, for a 3mm BiBO crystal cut for Type-I SPDC, with a pump wavelength of $775$nm, the temporal correlation width $\sigma^{(PM)}_{(t_{1}-t_{2})}$ is predicted to be approximately $4.0$fs, nearly two orders of magnitude below the width for a typical Type-II source. With a narrowband pump spectrum, the ratio between $\sigma_{(t_{1}+t_{2})}$ and $\sigma_{(t_{1}-t_{2})}$ can be as large as $10^{9}$, a degree of temporal correlation outstripping any spatial correlations discussed thus far. However, even with negligible jitter times and noise in photon counting apparatuses, these temporal correlations are not accessible unless the experimental setup used to measure these correlations accepts a relatively large range of frequencies; the degree of measurable temporal correlations is limited by the frequency-time Heisenberg relation \cite{mandel1995optical}. In addition, due to dispersion in optical fiber, the time correlation width measured by ideal photon detectors may still be significantly larger than what could be measured exiting the crystal. Indeed, just as there is a simple formula for the temporal spreading of Gaussian pulses due to dispersion, one can show that the final correlation width $\sigma_{(t_{1}-t_{2})_{(f)}}$ is related to the initial correlation width at the source $\sigma_{(t_{1}-t_{2})_{(i)}}$ in a similar way, i.e.,:
\begin{align}
\sigma_{(t_{1}-t_{2})_{(f)}}^{2}&\approx \sigma_{(t_{1}-t_{2})_{(i)}}^{2} +\ell_{M}^{2}\kappa_{M}^{2}\sigma_{(\omega_{1}-\omega_{2})}^{2},\\
\sigma_{(\omega_{1}-\omega_{2})}^{\text{type-I}}&\approx \sqrt{\frac{6}{L_{z}\kappa_{M}}}\nn\\
\sigma_{(\omega_{1}-\omega_{2})}^{\text{type-II}}&\approx \frac{2\pi c}{L_{z}\Delta n_{g}}\nn
\end{align}
where $\kappa_{M}$ is the group velocity dispersion constant for the medium, and $\ell_{M}$ is the length of the medium (e.g., an optical fiber). Furthermore, in type-II SPDC, the variance of $(\omega_{1}-\omega_{2})$ diverges, so we approximate $\sigma_{(\omega_{1}-\omega_{2})}$ as the half-width at half-maximum of the Sinc function describing the statistics of $(\omega_{1}-\omega_{2})$. 

As an example of the significance of dispersion on the measurement of time correlations, SMF-28 optical fiber at 1550nm has a group velocity dispersion of about $2.3\times10^{-26}s^{2}m^{-1}$, which means that while a time correlation width of $4.0$ fs of photon pairs exiting a nonlinear crystal is possible, this width is broadened by a factor of $\sqrt{2}$ after a distance of just $0.7$mm. Over longer distances, it approximately spreads by $5.8$ps for every meter of propagation in fiber. This dispersion can be compensated by propagating through an appropriate length of another medium of opposite group velocity dispersion (e.g., dispersion compensating optical fiber).

Using the conditional uncertainty relation(s) \eqref{CondUncRel}, we can study the tradeoff between the narrowness of the frequency spectrum of the signal or idler photon given by the standard deviation $\sigma_{\omega_{1}}$, and the temporal correlation width $\sigma_{(t_{1}-t_{2})}$. Using the same steps leading to \eqref{CondUncRel}, we find
\begin{equation}
\sigma_{\omega_{1}}\sigma_{(t_{1}-t_{2})}\geq\frac{1}{2},
\end{equation}
This is understood because the standard deviation is invariant to constant shifts, and that conditioning on average never increases the variance. Thus, we know $\sigma_{(t_{1}-t_{2})}\geq \sigma_{((t_{1}-t_{2})|t_{2})}=\sigma_{(t_{1}|t_{2})}$. Consider that a typical source of down-conversion has a frequency width $\sigma_{\omega_{1}}$ of the order $2\times 10^{14}$ (radians per second) \cite{BaekSPDCspectrumPRS2008}. This implies that the smallest possible time correlation width $\sigma_{(t_{1}-t_{2})}$ for this source is  of the order of $4$ femtoseconds. However, if we perform frequency filtering to look at the nearly degenerate part of the biphoton frequency spectrum, we increase the minimum resolvable time correlation width by a factor inversely related to the fraction of frequencies allowed to pass through the filter. If we consider a $2$nm filter centered at $1550$nm making $\sigma_{\omega_{1}} \approx 7.8\times10^{11}$ (radians per second), then the minimum resolvable value of $\sigma_{(t_{1}-t_{2})}$ would be about $6\times10^{2}$fs, more than two orders of magnitude wider that what is achievable without filtering.

\bibliography{EPRbib15}

\end{document}